# Nitrogen vacancy cluster-mediated ferroelectric domain growth in wurtzite AlN

Zijian Zhou[1], Kan-Hao Xue[1,*], Ziqi Pang[1,2], Heng Yu[1], Shanzhong Xie[1], Hang Liu[2,3,*], Haiming Qin[4], Yi Tong[4], Bowen Su[2,3], Yi Yang[2,3], Tian-Ling Ren[2,3*], and Xiangshui Miao[1]

[1]School of Integrated Circuits, Huazhong university of Science and Technology, Wuhan 430074, China

[2]School of Integrated Circuits, Tsinghua University, Beijing 100084, China

[3]Beijing National Research Center for Information Science and Technology, Tsinghua University, Beijing 100084, China

[4]Suzhou Laboratory, Suzhou 215123, China

*Corresponding authors: xkh@hust.edu.cn (K.-H. X.), liu_hang@mail.tsinghua.edu.cn (H. L.), rentl@tsinghua.edu.cn (T.-L. Ren)

## Abstract

Aluminum nitride-based ferroelectrics combine high remanent polarization, thermal stability, weak size effects, a wide bandgap and back-end process compatibility, making them promising for high density non-volatile memory. However, their high coercive field ($E_c$) and unclear polarization switching mechanism limit further applications. Theoretical predictions indicate preferential reversed domain growth along the polar axis, whereas experiments show nearly isotropic growth or preferential growth perpendicular to this axis. Here, we combine density functional theory, Deep Potential molecular dynamics and experimental characterization to determine how nitrogen vacancies ($V_N$) regulate domain growth and $E_c$. In ideal crystal, strong lattice coupling along the polar axis drives preferential domain growth in this direction. $V_N$ locally weakens this coupling and promotes polarization nucleation. The spatial distribution of $V_{\mathrm{N}}$ and the temporal sequence of vacancy induced switching jointly determine the direction and morphology of reversed domain growth. We further show that scandium doping promotes $V_{\mathrm{N}}$ formation, thereby enhancing defect mediated nucleation and reducing $E_{\mathrm{c}}$. These findings establish a multiscale mechanism governing thin-film domain-growth kinetics and macroscopic $E_{\mathrm{c}}$, reconciling experimental observations with ideal crystal simulations and providing defect engineering principles for low-voltage, reliable and scalable ferroelectric devices.

## Main

Wurtzite aluminium nitride (AlN) has long been regarded as a technologically mature piezoelectric material[1–3]. Its robust piezoelectricity, thermal stability and wide bandgap have enabled broad use in acoustic[4], electromechanical[5,6] and optoelectronic[7] devices. This conventional view changed fundamentally, nevertheless, after Fichtner et. al. demonstrated reversible ferroelectric polarization switching in scandium (Sc)-doped AlN[8]. AlN-based ferroelectrics, represented by $Al_{1-x}Sc_xN$, offer several key advantages for device applications[9,10]. Their polarization remains stable after 1100°C thermal annealing[11,12]. They also exhibit thickness scaling potential down to 5-20 nm[13,14] and a high remanent polarization[15,16] typically over $100\ \mu C/cm^2$. These properties provide a large readout window for high density non-volatile memory[17,18]. In addition, their bandgap of approximately $4-6$ eV[19] and process compatibility below 400°C are particularly attractive for back-end-of-line integration[20,21].

Despite rapid progress in device development, the microscopic mechanism of polarization switching in AlN-based ferroelectrics remains unsettled. Existing first principles studies mainly explain the Sc-induced coercive field ($E_c$) reduction in terms of the intrinsic energy landscape of ideal crystals[22]. In this picture, Sc modifies local bonding and lattice ionicity, thereby lowering the switching barrier[23]. However, this theory cannot capture the defect and charge effects commonly found in real thin films, such as preferential nucleation at interfaces[24,25] and the formation of interfacial dead layers[26]. Experiments have repeatedly shown that nitrogen vacancies ($V_N$) affect switching in AlN-based ferroelectrics[26,27]. However, their role in polarization switching and their possible contribution to $E_c$ reduction remain unclear at the atomic scale. A central question is therefore how strongly switching is governed by the intrinsic barrier or by defects in real materials.

This knowledge gap is particularly evident in domain growth kinetics, as it is widely accepted that the polarization switching in doped AlN is triggered by domain wall (DW) motion[28]. Advances in machine learning potentials now permit simulations of nucleation and DW propagation at experimentally relevant length scale within short timescales[29]. However, theoretical simulations and experiments yield different preferred directions of domain growth. Calderon et. al. observed nearly isotropic growth in boron-doped AlN by imaging at Ångström resolution[30]. Preferential lateral growth of reversed domains has also been reported experimentally[24]. A lateral direction denotes an in-plane direction perpendicular to the polar [001] direction. By contrast, Behrendt et al. employed a machine-learned force field to simulate AlN[31], whereas Huang et al. used a Deep Potential model for Sc-doped AlN[32]. Both simulations showed rapid penetration of a newly nucleated reversed domain along [001] ($z-$direction, longitudinal direction), followed by column-by-column (CBC) lateral growth. This behavior constitutes a pronounced longitudinal-first growth mode. Experimental isotropic or lateral-first growth therefore directly contradicts the reproducible longitudinal-first growth in simulations to ideal crystals. Ideal lattice models appear to be insufficient to describe domain evolution in real AlN-based ferroelectric films. Point defects, which are unavoidable in experimental specimens, may connect these two pictures.

Here, we combine theoretical simulation and experimental characterization to identify $V_{\mathrm{N}}$ as a key microscopic factor governing both the domain growth mode and the $E_{\mathrm{c}}$ of AlN-based ferroelectrics. Our molecular dynamics simulations reproduce growth modes spanning longitudinal-first propagation to lateral growth and reveal how $V_{\mathrm{N}}$ modifies local DW structure and propagation kinetics. We further resolve the atomic-scale mechanism by which $V_{\mathrm{N}}$ changes the domain growth mode. The Sc-induced $E_{\mathrm{c}}$ reduction cannot be attributed solely to changes in the intrinsic energy landscape of the ideal crystal. Rather, Sc also thermodynamically promotes $V_{\mathrm{N}}$ formation, opening another route to $E_{\mathrm{c}}$ reduction through defect mediated polarization switching. This mechanism further provides a unified microscopic picture of charge injection control over $E_{\mathrm{c}}$ in Sc-doped AlN. This work also offers a defect-mediated framework that reconciles experiments with ideal crystal simulations of AlN-based ferroelectrics.

## Results

*Domain growth molecular dynamics simulation*

Wurtzite AlN has a stable, uniform polarization orientation along $z-$direction, rendering its polarization switching strongly cooperative along this direction. As shown in **Fig. 1a** and **Supplementary Fig. 1a**, an artificially introduced lateral DW is unstable in pristine AlN. The tentative DW is perpendicular to $z-$direction and divides the system into upper and lower regions with opposite polarizations. After structural relaxation, the DW disappears and the system returns to a single polarization state. Climbing-image nudged elastic band (CI-NEB) calculations on a $5 \times 1 \times 5$ AlN supercell further support this result (**Supplementary Fig. 2**). Hence, layer-by-layer (LBL) DW propagation along $z-$direction cannot be sustained in the ideal lattice. The machine learning potential trained on density functional theory (DFT) data retains this physical behavior. Macroscopic domain growth therefore proceeds mainly in a CBC manner, rather than through stable LBL DW propagation. Introducing $V_{\mathrm{N}}$, however, breaks this strict cooperativity along $z-$direction. With just one $V_{\mathrm{N}}$ in the $5 \times 1 \times 5$ supercell, the $V_{\mathrm{N}}$ locally stabilizes a reversed polarization region that is otherwise unstable. The structural distortion around $V_{\mathrm{N}}$ effectively separates regions with opposite polarization orientations (**Supplementary Fig. 1b**). Thus, $V_{\mathrm{N}}$ provides a local degree of freedom for DW formation and propagation along the $z-$direction. DFT calculations further showed that $V_{\mathrm{N}}$ does not favor a completely random distribution (**Fig. 1b**). In a 142-atom model, clustered $V_{\mathrm{N}}$ had a total energy approximately 1.17 eV lower than dispersed $V_{\mathrm{N}}$. Based on this result, we constructed a thin film AlN model measuring approximately $10.8 \times 7.8 \times 5.2$ $\mathrm{nm}^3$ and containing $4 \times 10^4$ atoms. A cluster of three $V_{\mathrm{N}}$ was placed at its center (**Fig. 1c**).

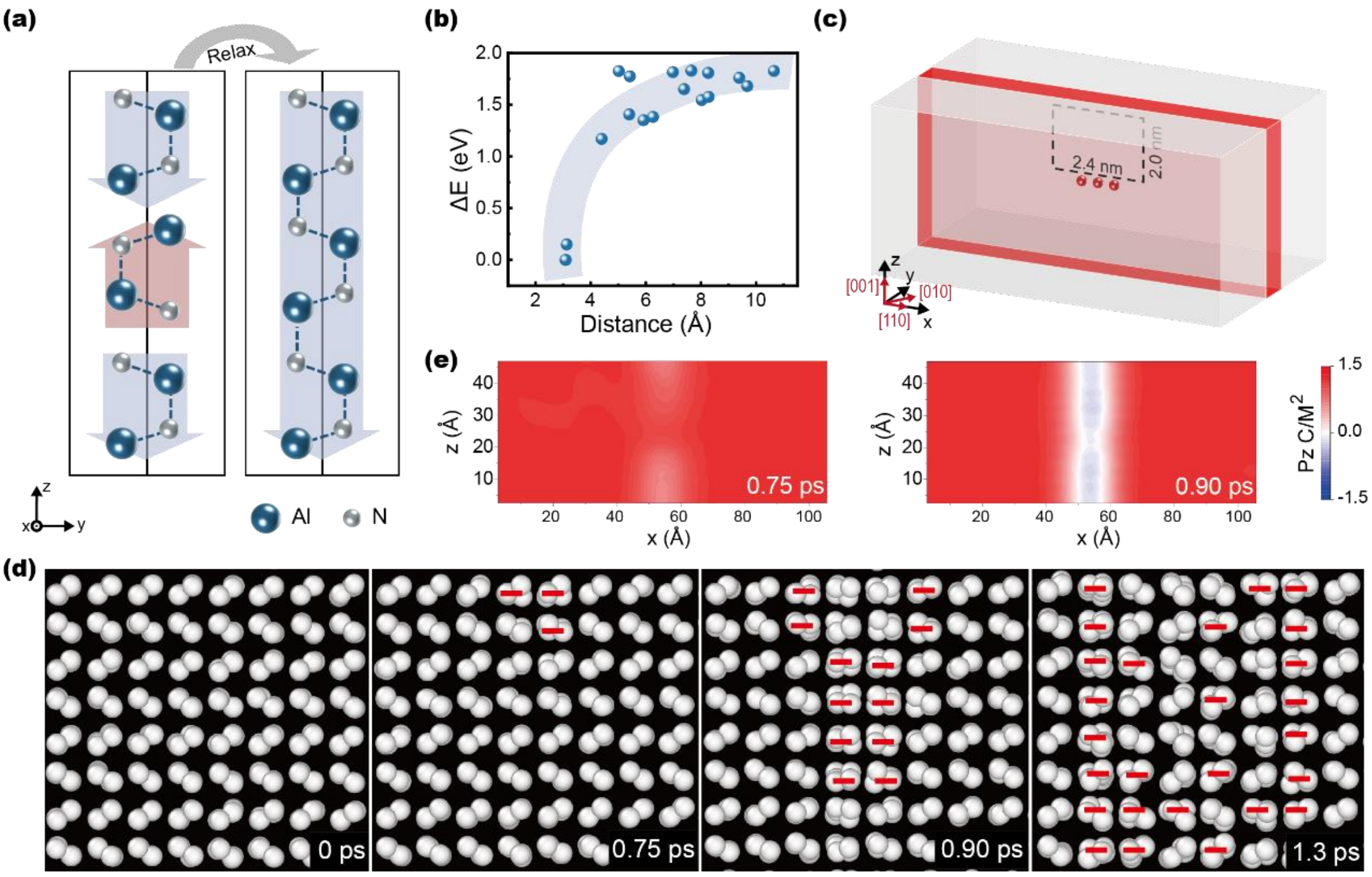

**Figure 1 | Cluster-$V_N$-mediated domain nucleation and growth in AlN. a,** Schematics of an artificially inserted DW before and after relaxation; the full supercell structure is shown in **Supplementary Fig. 1a. b,** Total energy of AlN as a function of the separation between $V_N$. **c,** Schematic of the cluster $V_N$ used in the MD simulation. **d,** Atomic structures projected along the $y$ −direction at different times. **e,** Polarization distributions at different times. **d** shows the black-boxed region in **c**, whereas **e** shows the red-plane slice.

We trained a deep potential (DP) model for molecular dynamics simulations using DeePMD-kit and Deep Generator. The training set extensively sampled configurations associated with polarization switching, nucleation, DWs and transition states. The dataset was also explicitly enriched with $V_N$-containing structures to describe the defect environments. The DP model accurately predicts local atomic environments and mechanical properties. Detailed validation is provided in **Supplementary Note 1**. Under a constant external field of 22 MV/cm, within a 2.4 nm × 2.0 nm observation window above the $V_N$ cluster, a localized isotropic domain expansion process was observed along the $y$ −direction, exhibiting a morphology similar to that reported in Ref. 25. The red curve in **Fig. 1d** marks the instantaneous DW position. A local reversed domain emerged at approximately 0.75 ps and then expanded simultaneously in longitudinal and lateral directions. This apparently isotropic growth does not imply an intrinsically isotropic DW propagation mechanism in AlN. Upon expanding the observation scale, as shown in **Fig. 1e**, the reversed domain is found to nucleate near $V_N$ and preferentially propagate forward along the $z$ direction. The apparent lateral growth observed in **Fig. 1d** results from the pointed DW at the domain front passing through the local observation window. The phenomenon was reproducible over fields of 22-27 MV/cm. This reproducibility excludes any incidental morphology caused by a single thermal fluctuation (**Supplementary Fig. 3**).

Experimentally, Mo/$Al_{0.9}Sc_{0.1}N$/Mo ferroelectric capacitors were fabricated through physical vapor deposition. The spontaneous polarization exceeds 150 μC/cm$^2$ as demonstrated in **Fig. 2a**. On Mo bottom electrode, the 50 nm-thick $Al_{0.9}Sc_{0.1}N$ film grew solely along the polar $c$-axis, supported by the X-ray diffraction pattern shown in **Fig. 2b**. To explore the spatial distribution of N vacancies along the polarization direction, **Fig. 2c** presents Al-$L_{2,3}$ edge EELS spectra acquired from three consecutive nanoscale regions in the upper, middle and lower parts of the $Al_{0.9}Sc_{0.1}N$ film interior, far from the interface. The spectra from all three regions nearly overlap in both line shape and peak position, exhibiting well-resolved and symmetric doublet features with no appreciable variation in their relative intensities. These results indicate a uniform local chemical environment throughout the film interior, consistent with the rapid recovery of a highly crystalline and compositionally homogeneous wurtzite bulk phase away from the interface. **Fig. 2d** compares the Al-$L_{2,3}$ edge fine-structure from the bulk dielectric region of the AlScN film and the interfacial region adjacent to the bottom Mo electrode. In the bulk dielectric region, the Al-$L_{2,3}$ edge exhibits a characteristic doublet fine structure associated with the unoccupied local electronic states of tetrahedrally coordinated Al–N4 units in the ideal wurtzite lattice. By contrast, pronounced peak broadening and attenuation of the fine-structure features are observed near the bottom Mo/$Al_{0.9}Sc_{0.1}N$ interface. The degradation of this fine structure indicates a substantial loss of local nitrogen coordination at the interface, providing spectroscopic evidence for $V_N$ enrichment in this region.

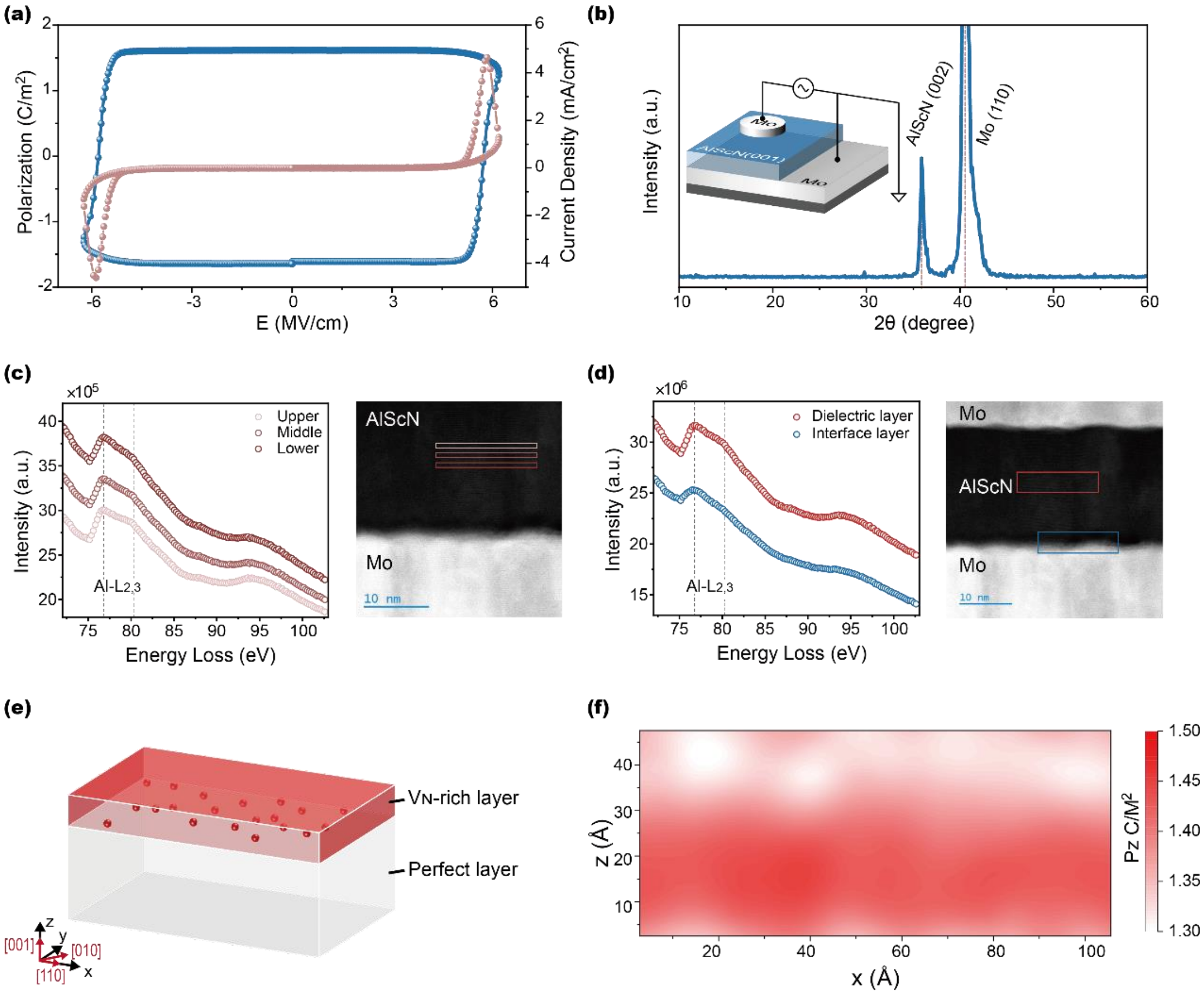

**Figure 2 | Domain growth from a $V_N$-rich layer. a,** P–E and J–E curves of the $Al_{0.9}Sc_{0.1}N$ film obtained from PUND measurements. **b,** X-ray diffraction pattern of the $Al_{0.9}Sc_{0.1}N$ film and schematic of the Mo/AlScN/Mo capacitor. **c,** EELS spectra of Al-$L_{2,3}$ edge acquired at different depths within the $Al_{0.9}Sc_{0.1}N$ film. **d,** EELS spectra of Al-$L_{2,3}$ edge from the dielectric layer and interface layer regions of the $Al_{0.9}Sc_{0.1}N$ film. **e,** Schematic of the $V_N$-rich layer used in the MD simulation. **f,** Polarization distribution in the $x-z$ plane projected along the $y$ −direction.

To model this situation, we randomly introduced 300 $V_N$ within an approximately two unit cell thick region, about 1 nm, at the film surface (see **Fig. 2e**). Subsequently, a constant 21 MV/cm external field was applied. **Fig. 2f** illustrates the local polarization projected onto the $x-z$ plane. White areas represent unit cells that have been switched to the reversed polarization, and the corresponding atomic structures are shown in

**Supplementary Fig. 4**. The behavior differed markedly from that with cluster $V_{\mathrm{N}}$. Shortly after field application, a reversed domain formed in the $V_{\mathrm{N}}$-rich region and grew rapidly in lateral directions. It eventually spanned the entire lateral dimension of the film. The reversed domain position closely coincided with the prescribed $V_{\mathrm{N}}$-rich layer.

*Mechanistic Origin of the Domain Growth Mode*

To identify the common mechanism underlying both CBC and LBL growth modes, we first examined how $V_N$ selects the nucleation site of a reversed domain. Nine $V_N$ were randomly introduced into the model, and $20$ independent simulations were performed under an electric field of 21 MV/cm (**Fig. 3a** and **Supplementary Fig. 5**). It follows that the nucleation sites of reversed domains were strongly correlated with the $V_N$ positions. Regions near $V_N$ showed a substantially higher nucleation probability than pristine lattice regions. At a fixed evolution time, a larger local reversed polarization area generally indicates earlier nucleation. Regions with denser $V_N$ distributions had larger reversed polarization areas, indicating earlier nucleation at higher local vacancy concentrations. This trend agrees with the static energy barrier calculations. CI-NEB calculations gave a local nucleation barrier of approximately 0.77eV per unit cell near $V_N$. This value is approximately 0.31 eV lower than that in pristine AlN (**Fig. 3b**). Thus, $V_N$ reshapes the local energy landscape from the onset of switching and makes its neighbourhood a preferred nucleation region.

The CI-NEB pathway structures further reveal the domain growth mode after $V_N$-induced local nucleation. **Fig. 3c** shows $z-$direction growth near $V_N$ in images 1–4 and lateral growth in images 4–6. During switching, the alternating long and short Al–N bonds along $z-$direction were progressively disrupted. The local Al–N distances near $V_N$ became more uniform, forming a ring of 10 members (highlighted in orange color), resembling the non-polar $P6_3/mmc$ phase. As switching proceeded, the reversed polarization region extended along $z-$ direction through LBL DW propagation. It eventually yielded a columnar reversed domain spanning the crystal. Upon entering $V_N$-free regions, domain growth reverted to the CBC mode. The principal effect of $V_N$ is therefore not a direct shift of the intrinsic domain growth direction from longitudinal to lateral. Instead, it only transforms polarization switching from a strongly cooperative process in the ideal lattice into preferential local nucleation centered on defects.

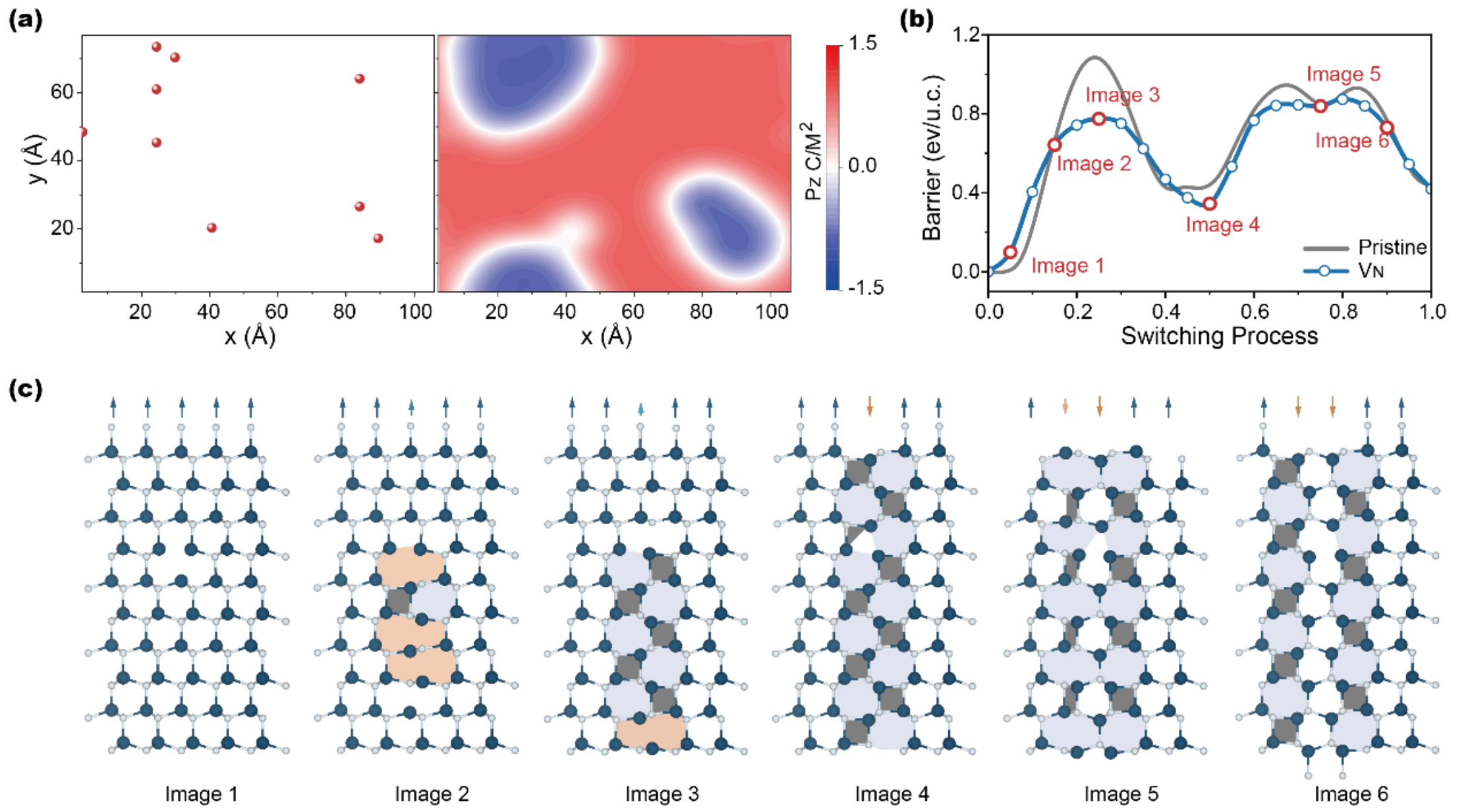

**Figure 3 | Preferential nucleation and atomistic switching near $V_N$. a,** $V_N$ distribution projected and the polarization distribution at nucleation along $z$ −direction. **b,** CI-NEB barrier for nucleation and domain growth in a $5 \times 1 \times 5$ AlN supercell containing one $V_N$. **c,** Atomic structures corresponding to **b**. Orange regions mark the domain wall during growth along $z$ −direction; blue and grey regions form the DW during lateral growth.

This mechanism unifies the two apparently distinct domain growth modes described above. For a local $V_N$ cluster, individual sites experience distinct local defect environments, subject to nucleation at different times (**Fig. 4a**). Switching first occurs at the cluster center, where the $V_N$ concentration is higher and the barrier is lower. Neighboring $V_N$ sites are then activated sequentially. Each nucleation event first triggers LBL growth along $z$ −direction. Meanwhile, a neighboring $V_N$ nucleates before the preceding reversed domain fully spans the crystal. Reversed domains with various onset times therefore emerge along $z$ − direction, grow and coalesce continuously. Within a finite observation window, the temporally staggered nucleation and spatial overlap appear as simultaneous lateral and longitudinal DW propagation. This process produces the pseudo-isotropic domain growth observed in **Fig. 1d**. The same mechanism operates in an interfacial $V_N$-rich layer, but the preferred nucleation sites differ. Many $V_N$ lie at nearly the same film depth, creating numerous neighboring sites with low nucleation barriers. Under an external field, their neighborhoods switch successively or nearly synchronously within a short interval. These reversed regions rapidly connect in-plane and first generate a laterally spanning reversed domain. Switching then propagates along $z$ −direction from this nucleated interface into the film interior, where the $V_N$ concentration is lower. This sequence produces the lateral-first and then growth along $z$ −direction.

Preferential nucleation also directly affects $E_c$. Comparative simulations with varying random $V_N$ concentrations showed a marked reduction after introducing even a few vacancies. At a $V_N$ concentration of approximately 1%, $E_c$ decreased from 30.84 MV/cm in pristine AlN to 21.75 MV/cm, corresponding to ~30.1% reduction (**Fig. 4b**).

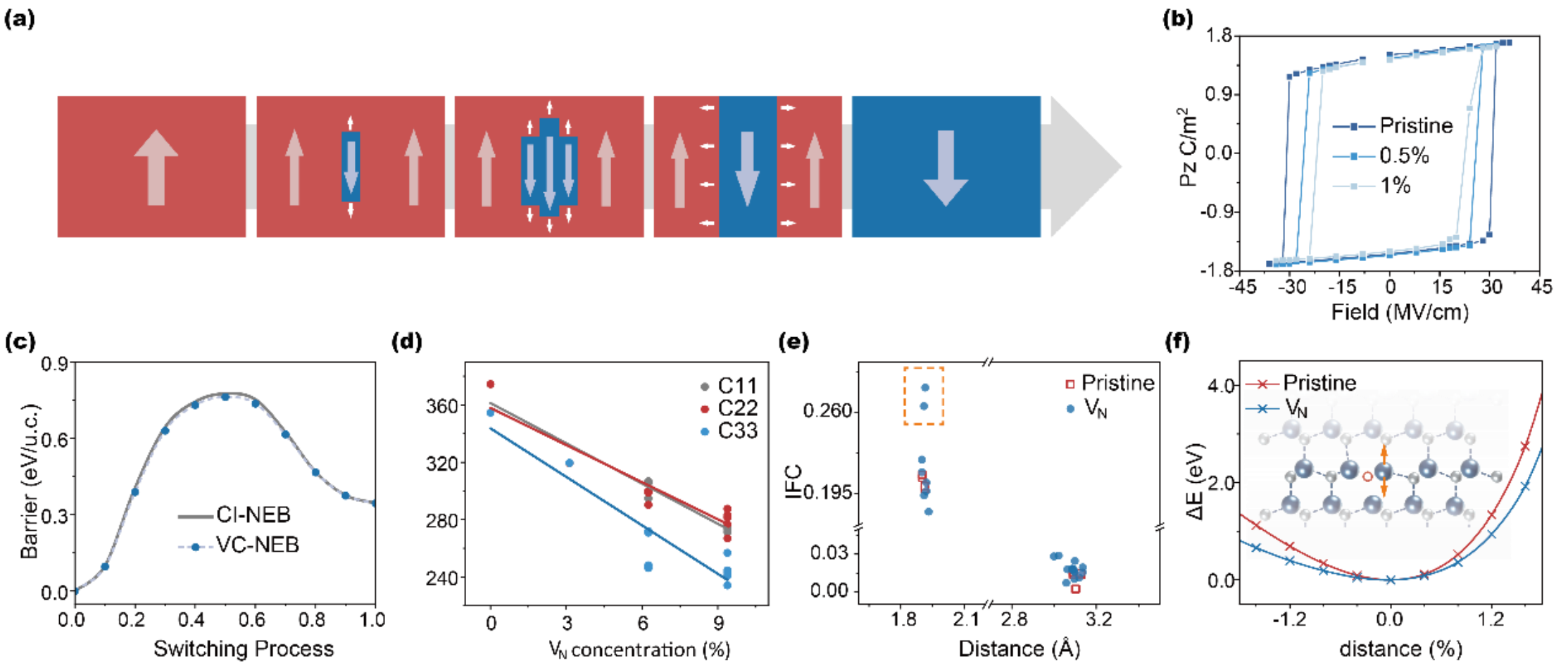

**Figure 4 | Mechanistic origin and $E_c$ consequence of $V_N$-mediated nucleation. a,** Schematic of sequential nucleation induced by a $V_N$ cluster. **b,** MD simulated hysteresis loops for pristine AlN and AlN with randomly distributed $V_N$ at varying concentrations. **c,** CI-NEB and SS-NEB barriers for nucleation in a $5 \times 1 \times 5$ AlN supercell containing one $V_N$. **d,** Elastic constants $C_{11}$, $C_{22}$ and $C_{33}$ as functions of $V_N$ concentration, with linear fits. **e,** The normalized interatomic force constants in pristine and $V_N$-containing AlN. **f,** Local displacement potential of the Al atom nearest to $V_N$ compared with that of an Al atom in pristine AlN. The horizontal axis represents the displacement of the Al atom from its equilibrium position, normalized by the $z$ −axis lattice parameter of the unit cell.

The preceding results establish a direct link between the spatial distribution of $V_N$, preferential nucleation and the domain growth morphology. A fundamental question remains: why does reversed domain nucleate near $V_{\mathrm{N}}$ with a much lower barrier? To identify its microscopic origin, we calculated local nucleation pathways in $V_{\mathrm{N}}$-containing AlN using CI-NEB and solid-state nudged elastic band (SS-NEB). Both methods yielded nearly identical barriers (**Fig. 4c**). Thus, the barrier reduction does not mainly arise from global cell-volume relaxation during the transition. Nor does it reflect numerical differences between the pathway search methods. Instead, it originates from an intrinsic change in local bonding and interatomic interactions near the defect.

We first analyzed the elastic response of $P6_3mc$ AlN at various $V_{\mathrm{N}}$ concentrations. Multiple defect configurations were constructed and fitted at each concentration to reduce statistical bias from random $V_{\mathrm{N}}$ arrangements. The overall elastic constants decreased progressively upon increasing $V_{\mathrm{N}}$ concentration (**Fig. 4d**). Notably, $C_{33}$ decreased substantially faster than $C_{11}$ and $C_{22}$. Such pronounced anisotropic softening indicates that $V_{\mathrm{N}}$ preferentially weakens the lattice stiffness along the $z-$direction. This effect is particularly important for polarization switching in wurtzite AlN. Switching requires relative cation and anion displacements along $z-$direction. In pristine AlN, strong restoring forces render these displacements highly cooperative along $z-$direction. As $C_{33}$ decreases, the collective restoring force along the polar direction weakens substantially. The cooperative switching that initially requires multiple unit cells consequently starts to decouple locally.

Analysis of the normalized interatomic force constants reveals the highly localized nature of this softening. As shown in **Fig. 4e**, the most pronounced changes appear on short range interactions between the central $V_{\mathrm{N}}$ and the closest Al atoms, as highlighted by the orange dashed box. Longer range interactions and Al atoms far from the defect retain responses close to those in pristine AlN. Thus, rather than uniformly weakening the lattice, $V_{\mathrm{N}}$ locally reconstructs the restoring force network around the defect. Importantly, an increase in the normalized nearest neighbor interatomic force constant does not indicate stronger confinement of the surrounding Al atoms. Removing one N atom disrupts the multidirectional restoring force network provided by the complete coordination environment. The residual restoring force is redistributed among a few remaining bonds. An individual short range interaction may therefore get strengthened, even though the total local potential becomes shallower. Local displacement calculations along $z$ −directionconfirm this interpretation directly. The energy displacement curve of an Al atom adjacent to $V_{\mathrm{N}}$ is flatter than that in pristine AlN (**Fig. 4f**), reflecting a lower local restoring force and reduced potential-well curvature.

The physical origin of the lower nucleation barrier can therefore be summarized as follows. N removal disrupts the intact local force network and preferentially softens the restoring forces along $z$ −direction that directly govern polarization switching. Atoms near the defect consequently become partially decoupled from the highly cooperative lattice constraint. Under an external electric field, these softened regions enter the non-polar transition configuration first, followed a complete reversal into the opposite polarization.

*Sc Doping Enhances the $V_N$-Mediated Effect*

We calculated the formation energies of $V_\mathrm{N}$ at several N sites near Sc, showing that Sc lowered the vacancy formation energies at several sites under both N-rich and Al-rich conditions (**Fig. 5a**). This trend was not restricted to any specific atomic configuration, thus the presence of Sc promotes $V_\mathrm{N}$ formation. Experimental results also support this prediction. For films with different Sc concentrations, Al-$L_{2,3}$ edge EELS spectra were acquired from three consecutive regions in the upper, middle and lower parts of the film interior, away from the interface. The low-Sc film (x = 0.10) exhibits a clear and symmetric doublet fine structure, whereas the doublet becomes substantially less pronounced at higher Sc concentration (x = 0.18). This spectral evolution indicates increasing stoichiometric deviation in the high-Sc film, consistent with a higher concentration of $V_\mathrm{N}$ (**Fig. 5b**). CI-NEB calculations of switching pathways at different $V_\mathrm{N}$ sites in Sc-doped AlN gave a consistent picture. Most $V_\mathrm{N}$ sites further lowered the local switching barrier, with $V_\mathrm{N}$ at the N2 site reducing the barrier by 0.31 eV per unit cell, corresponding to a decrease of approximately 45% (**Fig. 5c**). We used the representative N2 site to calculate formation energies at different Fermi levels and charge states. **Fig. 5d** reveals that the thermodynamic stability of $V_\mathrm{N}$ is sensitive to the local charge environment. Carrier injection lowers its formation energy within a specific Fermi level range. Positively charged vacancies also further reduce the corresponding polarization switching barrier relative to neutral vacancies (see **Supplementary Fig. 6**).

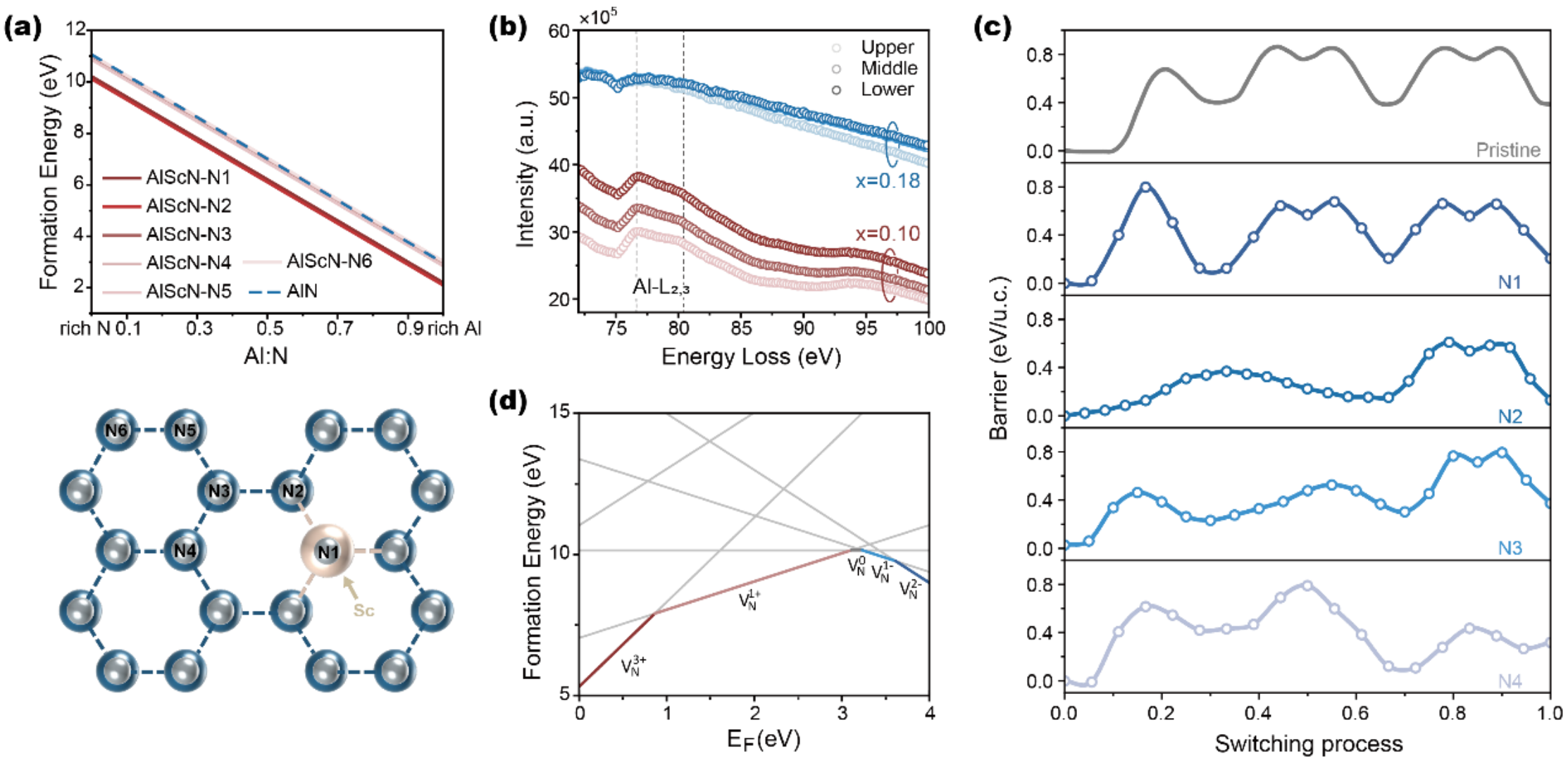

**Figure 5 | Effects of Sc and charge state on $V_N$ formation and switching. a,** Formation energies of $V_N$ at different sites in Sc-doped AlN. **b,** EELS spectra of Al-$L_{2,3}$ edge acquired at different depths within the $Al_{1-x}Sc_xN$ films with various $x$ composition. **c,** CI-NEB barriers associated with $V_N$ at different sites in Sc-doped AlN. The first peak corresponds to the nucleation barrier. **d,** Formation energy of charged $V_N$ at the N2 site as a function of Fermi level.

## Discussion

We have established a domain growth mechanism in AlN-based ferroelectrics governed by vacancy formation, local force network softening, preferential nucleation and defect distribution. The presence of $V_{\mathrm{N}}$ disrupts the strong cooperative constraint in AlN and enables preferential nucleation of reversed polarization near the defects. Following this local mechanism, spatial distributions of $V_{\mathrm{N}}$ then yield distinct macroscopic domain morphologies. Softening of the local restoring force network, particularly along the polar axis, is the physical origin of this behavior. Sc doping thermodynamically promotes the formation of $V_{\mathrm{N}}$, making this mechanism more active. The higher concentration and random distribution of Sc-induced $V_{\mathrm{N}}$ may also produce more diverse domain growth morphologies. The Sc-induced $E_{\mathrm{c}}$ reduction should therefore be understood as a combined consequence of intrinsic lattice and $V_{\mathrm{N}}$-mediated regulation. Our work demonstrates that $V_{\mathrm{N}}$ can play a pronounced beneficial role in ferroelectric AlN, providing a new perspective for the further development of wurtzite ferroelectric materials. Our work demonstrates the beneficial role of $V_{\mathrm{N}}$ in ferroelectric AlN and highlights defect engineering as a promising strategy for the design of wurtzite ferroelectric materials.

# Methods

## Device fabrication

The samples were fabricated using an ultrahigh-vacuum magnetron sputtering system equipped with high-purity aluminum (Al, 99.99%) and scandium (Sc, 99.9%) targets. The base pressure of the growth chamber was maintained below $5 \times 10^{-9}$ Torr. Before film deposition, the Si substrates were sequentially ultrasonically cleaned in acetone, isopropanol and deionized water, and subsequently dried under a high-purity $N_2$ flow. The bottom Mo electrode was deposited by electron-beam evaporation.

The ferroelectric $Al_{1-x}Sc_xN$ layer was then deposited by reactive co-sputtering from the Al and Sc targets. During deposition, the substrate temperature was maintained at 350 °C. The Al target was powered by a DC source, whereas the Sc target was driven by an AC source, with both powers set to 80 W. $N_2$ was introduced as the reactive sputtering gas, and the deposition was continued for 1 h. Finally, the top Mo electrode was deposited by electron-beam evaporation and patterned using a lift-off process.

## Material characterization

X-ray diffraction (XRD) patterns were acquired with a Bruker D8 Advance X-ray diffractometer. Cross-sectional transmission electron microscopy (TEM) samples were prepared using a Zeiss focused ion beam (FIB) system. TEM imaging, electron energy loss spectroscopy (EELS), and energy-dispersive X-ray spectroscopy (EDS) analyses were performed on a JEM-ARM300F2 transmission electron microscope. Electrical measurements were carried out using a semiconductor characterization system (Keithley 4200A-SCS), a semiconductor device analyzer (Keysight B1500A), and a high-voltage ferroelectric tester (Precision Multiferroic II).

## Deep potential molecular dynamics

The deep potential model was developed using DP-GEN[33] and DeePMD-kit[34]. During the concurrent learning procedure, the trust region for the model force deviation was set to $0.10–0.28\ eV/Å$. In each iteration, the model was trained for $6 \times 10^6$ steps to ensure sufficient optimization on the accumulated training dataset. The $se_a$ descriptor implemented in DeePMD-kit was used to represent the local atomic environments, with a cutoff radius of 6.0 Å and a smoothing onset radius of 1.0 Å. The embedding network consisted of three hidden layers containing 25, 50 and 120 neurons, respectively, whereas the fitting network comprised three hidden layers with 240 neurons each. Details of the loss function, weighting parameters and learning-rate schedule are provided in the **Supplementary Note 1**. All molecular dynamics simulations were performed using LAMMPS[35] with the trained DP potential.

### First Principles Calculations

All first principles calculations were performed with the Vienna Ab initio Simulation Package (VASP)[36,37] within density functional theory (DFT). We used the projector augmented-wave (PAW) method[38,39] and the Perdew-Burke-Ernzerhof (PBE) functional[40]. The valence electron configurations were 2s2p for N, 3s3p for Al, and 3p4s3d for Sc. A plane-wave cutoff energy of 500 $eV$ was used in all calculations. During structural relaxation, all ionic forces were converged below 0.01 $eV/Å$. Both climbing-image nudged elastic band[41] and solid-state nudged elastic band[42] calculations were performed with the assistance of the VTST tools, using an ionic force convergence criterion of 0.03 $eV/Å$. Detailed computational procedures for the Born effective charges, normalized traces of the interatomic force constant tensors, vacancy formation energies and other related quantities are provided in the **Supplementary Note 2**.

## Data availability

The data that support the plots within this paper and other findings of this study are available from the corresponding authors upon reasonable request.

## Acknowledgement

This work was supported by the National Key R&D Program of China under Grant No. 2023YFB4402302, the National Natural Science Foundation (92580209, U25A6021) of China and Original Research Projects of the School of Integrated Circuits of Tsinghua University under Grant 53120600125.

## Author contributions

Z. Z. and K.-H. X. planned the study and performed the simulations. H. L. designed the experiments. K.-H. X., T.-L. R., and X. M. supervised the project. H. Q. and Y. T. grew the thin films, fabricated the devices, and performed EELS and XRD characterization. H. Y. and S. X. carried out the first-principles calculations. Z. P., B. S. and Y. Y. analyzed the electrical data. Z. Z., H. L., and K.-H. X. wrote the manuscript with input from all co-authors. All authors discussed the results and contributed to the final manuscript.

## Ethics declaration

### Competing interests

The authors declare no competing interests.

<u>Supplementary Information</u>

# Nitrogen vacancy cluster-mediated ferroelectric domain growth in wurtzite AlN

Zijian Zhou[1], Kan-Hao Xue[1,*], Ziqi Pang[1,2], Heng Yu[1], Shanzhong Xie[1], Hang Liu[2,3,*], Haiming Qin[4], Yi Tong[4], Bowen Su[2,3], Yi Yang[2,3], Tian-Ling Ren[2,3*], and Xiangshui Miao[1]

[1]School of Integrated Circuits, Huazhong university of Science and Technology, Wuhan 430074, China

[2]School of Integrated Circuits, Tsinghua University, Beijing 100084, China

[3]Beijing National Research Center for Information Science and Technology, Tsinghua University, Beijing 100084, China

[4]Suzhou Laboratory, Suzhou 215123, China

*Corresponding authors: xkh@hust.edu.cn (K.-H. X.), liu_hang@mail.tsinghua.edu.cn (H. L.), rentl@tsinghua.edu.cn (T.-L. Ren)

# CONTENTS

## Note 1. Deep Potential

To simulate polarization switching process in AlN thin films, we constructed in this work a Deep Potential (DP) model for AlN containing nitrogen vacancies ($V_N$) using the Deep Potential Generator (DP-GEN) and DeePMD-kit packages. DeePMD is a deep neural network based interatomic potential framework used for molecular dynamics (MD) simulations. In this framework, local atomic environment data obtained from first principles calculations are used as input. Through descriptor and fitting networks that preserve translational, rotational, and permutation symmetries among atoms of the same species, the total energy is expressed as the sum of atomic contributions, $E = \sum_i E_i$, and the potential energy surface and atomic forces are fitted accordingly.

### DP-GEN

DP-GEN is a concurrent active learning framework for constructing machine learning interatomic potentials. Its central idea is to obtain a highly accurate DP model with broad coverage of the target configurational space at relatively low first principles cost through a closed loop iterative procedure of "exploration-labeling-training". Specifically, DP-GEN first uses the potential obtained in the current iteration to explore the configurational space, through large scale MD sampling, thereby generating candidate structures. Representative configurations are then selected according to a model deviation criterion and labeled by first principles calculations to obtain their energies, atomic forces, and stresses. Finally, the newly labeled data are incorporated into the training set to update the DP model iteratively. This active learning workflow, reduces redundant labeling and improves coverage of complex local environments and nonequilibrium configurations, thereby combining first principles accuracy with the efficiency required for large scale atomistic simulations.

In this work, the trust interval for the force deviation was set to $0.10\text{–}0.28\ eV/\text{Å}$. Specifically, when the model prediction deviation fell within this range, the corresponding configuration was selected for first principles labeling in order to improve the representativeness of the training data and the generalization capability of the model. In addition, the number of training steps for each model in every iteration was set to $6 \times 10^6$ to ensure sufficient optimization of the DP model on the given dataset.

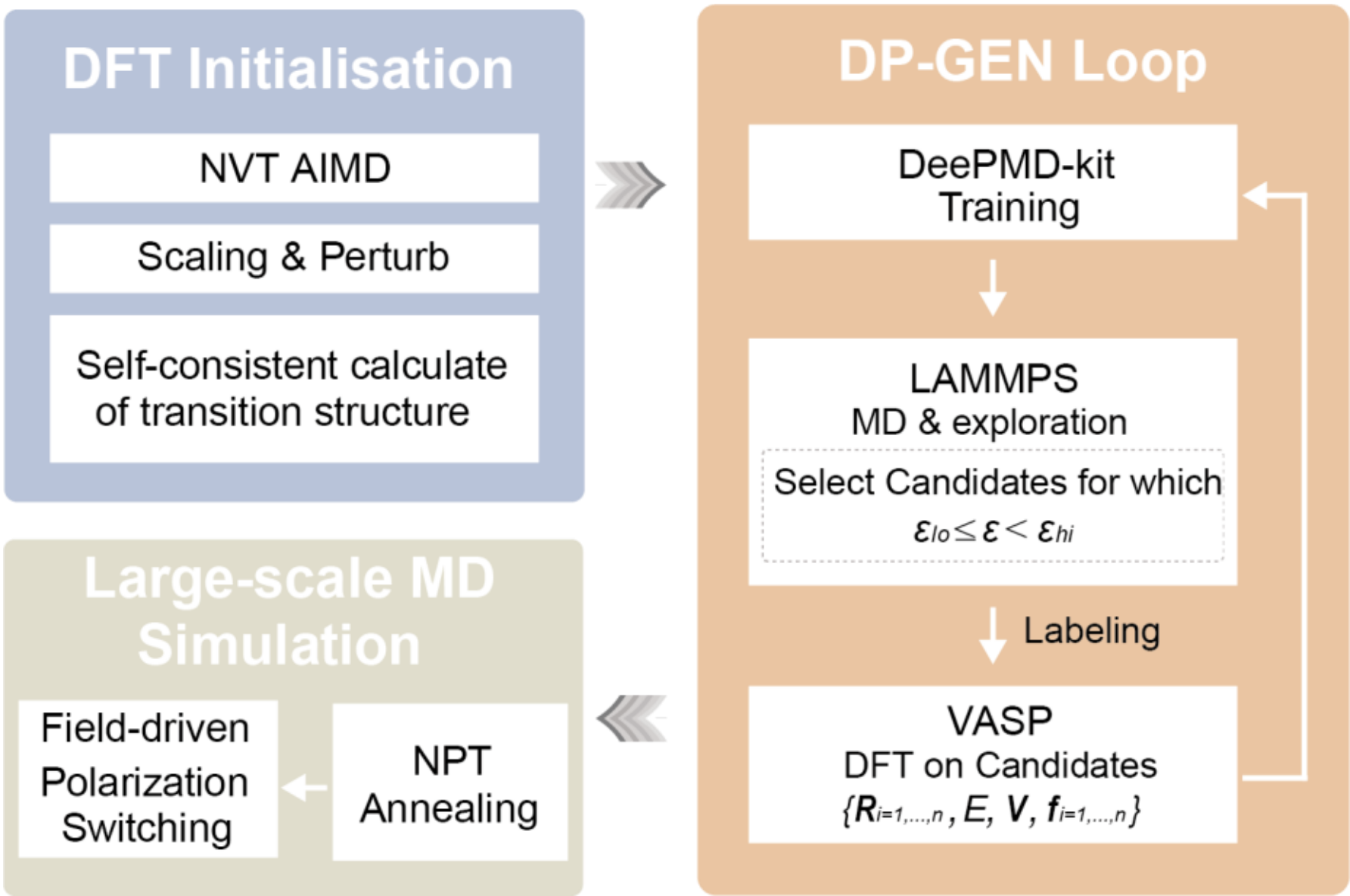

**Figure N1.** Workflow from DFT calculations to DP training and then to large scale MD simulations.

**Initial training database and exploration protocol**

The initial training dataset included the $P6_3mc$, $P6_3/mmc$, $\beta$ −BeO-like, $Fm3m$, $F\bar{4}3m$ phases of AlN, as shown in **Fig. N2**. The number of atoms was kept within the range of 128 to 216. To ensure a stable start of the training process, the initial dataset included perturbed structures of these phases together with AIMD data at 100 K, 300 K, and 800 K. During the exploration stage, a large number of configurations with randomly distributed $V_N$ defects across multiple phases were sampled to ensure accurate prediction of defect containing structures. Steady state and transition state configurations were both incorporated at this stage, including planar domain wall (DW), core DW, nucleation structures, and DW propagation configurations, so that the training data covered the atomic environments involved in the Nucleation and growth dynamics of inverted domains. The configurational space was sampled using NPT and NVT ensembles with thermostats at multiple temperatures. In total, approximately 2.03 million configurations were sampled throughout the exploration process.

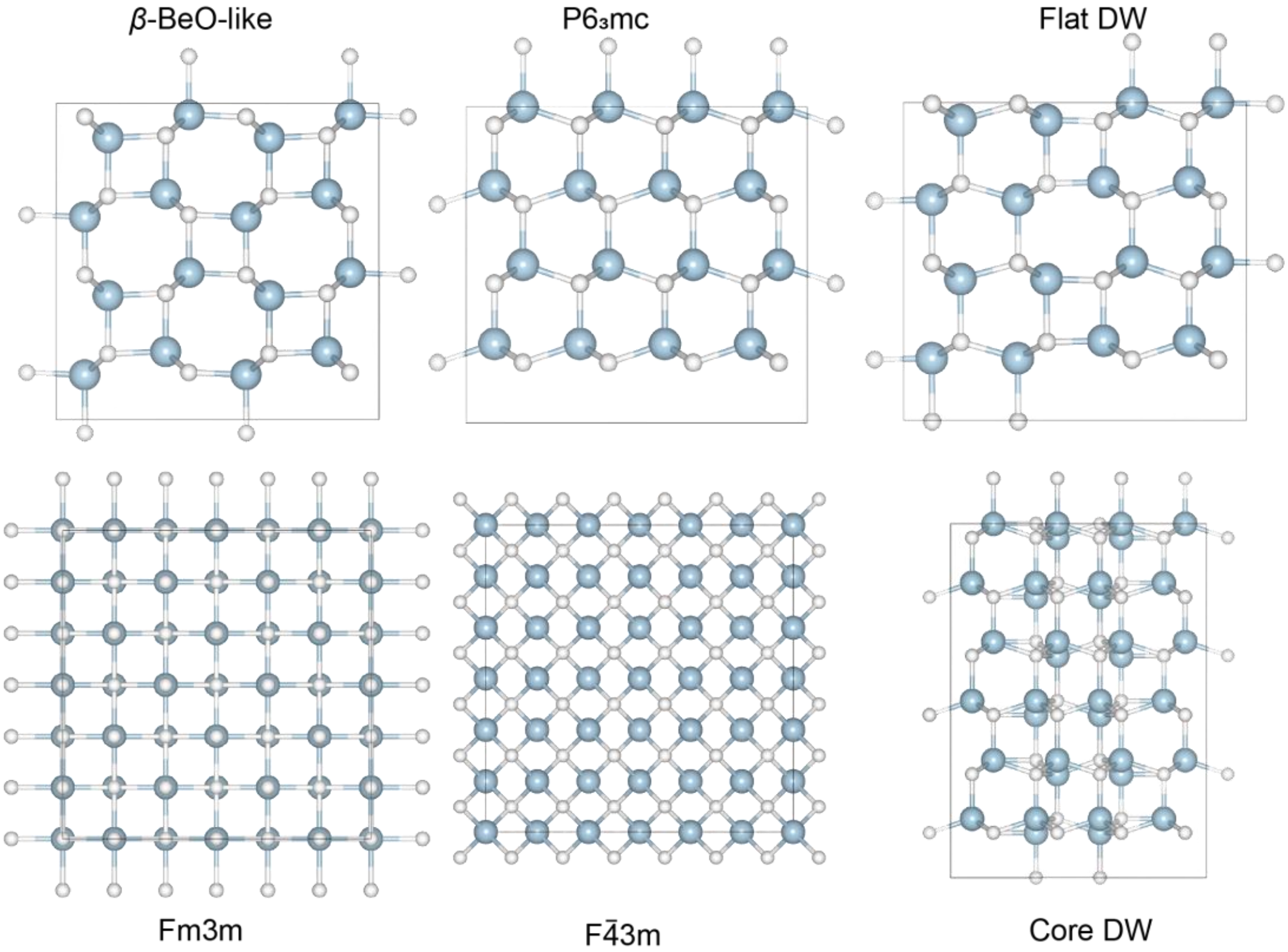

**Figure N2.** Structures used for training.

## Model parameters

The DP model was trained with DeePMD-kit. We used the *se_a* descriptor with a cutoff radius of 6.0 Å and a smoothing start radius of 1.0 Å. The embedding network architecture was [25, 50, 120], and the fitting network architecture was [240, 240, 240]. The loss function is defined as follows:

$$L(p_\epsilon, p_f, p_\xi) = p_\epsilon \Delta\epsilon^2 + \frac{p_f}{3N}\sum_i |\Delta F_i| + \frac{p_\xi}{9}\|\Delta\xi\|^2$$

here, the $\Delta$ terms denote the deviations between the DP predictions and the training data, $N$ is the number of atoms, $\epsilon$ and $F_i$ represent the atomic energy and atomic forces, respectively, and $\xi$ donates the virial tensor. The weighting factors $p_\epsilon$, $p_f$ and $p_\xi$ are adjusted during training by linear interpolation with the learning rate, that is:

$$p_\alpha(r) = p_\alpha^{limit} + (p_\alpha^{start} - p_\alpha^{limit})\frac{r}{r_0}$$

In this expression, $\alpha \in \{\epsilon, f, \xi\}$, where $r$ and $r_0$ denote the current and initial learning rates, respectively. The initial weighting parameters, $p_\epsilon^{start}$ , $p_f^{start}$ and $p_\xi^{start}$ ,were set to 0.001, 0.02, and 1000, respectively. The limiting weighting parameters, $p_\epsilon^{limit}$, $p_f^{limit}$ and $p_\xi^{limit}$ were all set to 1.0. An exponentially decaying learning rate schedule was employed for model training, with an initial learning rate of $1.0\times10^{-3}$ and a decay step of 20000.

**Model validation**

To ensure the reliability of the large scale atomistic simulations, the total energies and atomic forces predicted by the DP model were rigorously compared with those in the DFT dataset, thereby assessing the ability of the model to accurately describe local atomic environments accurately. As shown in **Fig. N3**, the gray line denotes perfect agreement between the DP and DFT values. The energy data lie almost entirely on this line, indicating excellent consistency, with more than 98.3% of the data exhibiting errors below 5 $meV/atom$. The three Cartesian components of the atomic forces are mainly concentrated in the range from -3 to 3 $eV/\text{Å}$. A small number of data points show errors approaching 1 $eV/\text{Å}$, which can be attributed to the inclusion of many unstable transition states from NEB calculations in the dataset. Nevertheless, after extensive subsequent training, the actual force errors remained very small: for $F_x$, $F_y$ and $F_z$, more than 92.2%, 91.6%, and 90.1% of the data, respectively, exhibit errors below 0.04 $eV/\text{Å}$. The excellent agreement in both energies and atomic forces over a broad range confirms that the developed potential can reproduce the potential energy surface of $V_N -$ AlN with high accuracy, thereby providing a reliable and computationally efficient tool for our MD simulations.

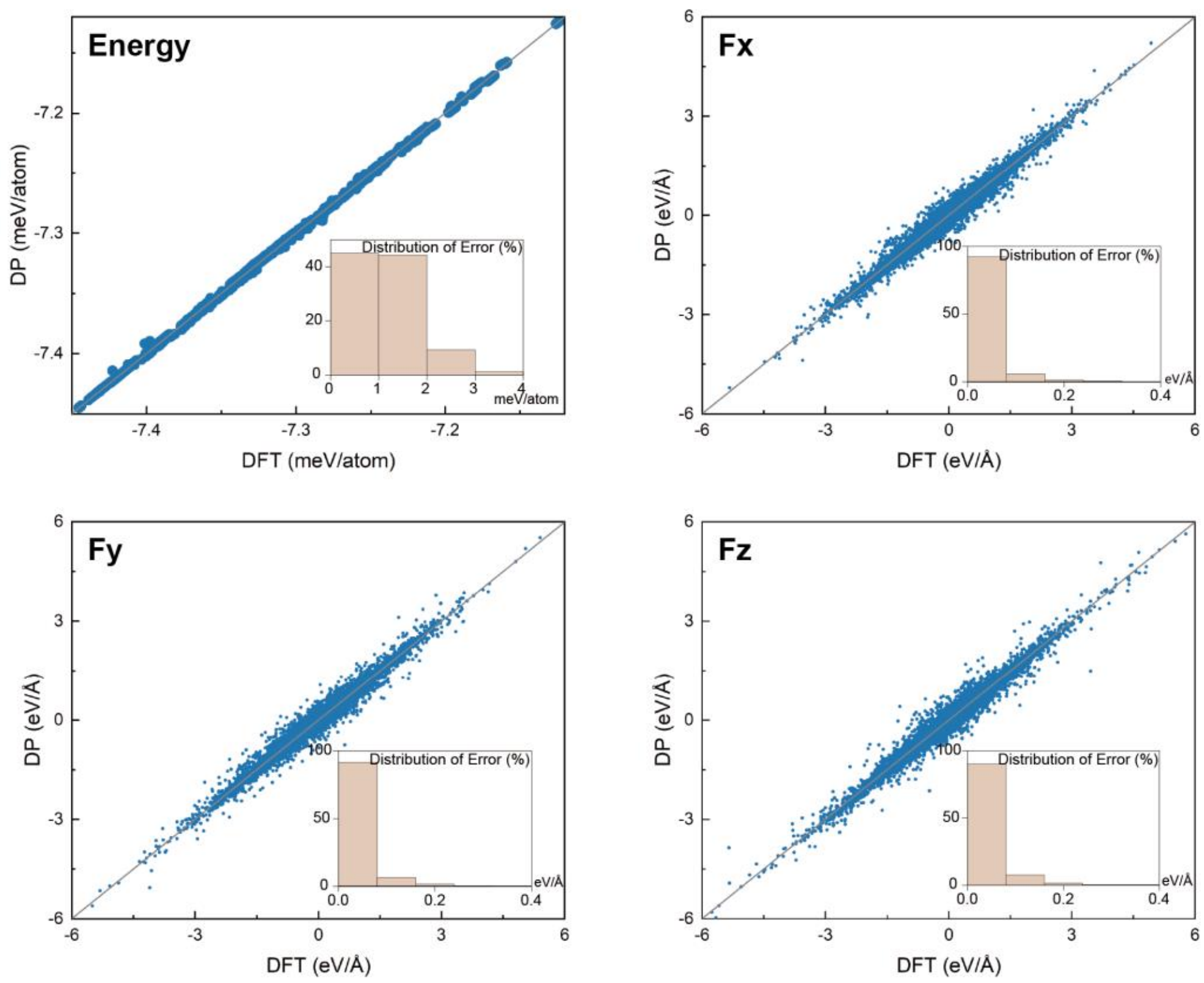

**Figure N3.** Comparison of atomic energies and forces predicted by the DP model and calculated by DFT.

To further validate the model at a more macroscopic level, we compared physical parameters of several phase structures calculated by DP and DFT. **Fig. N4** presents the elastic constants of the pristine AlN $P6_3mc$, $P6_3/mmc$ and $\beta$ −BeO-like phases, as well as the $P6_3mc$ phase containing one $V_N$, predicted by DP and calculated by DFT. Excellent agreement is achieved between the two methods. The corresponding numerical values of these elastic constants are listed in Table N1, which also includes comparisons of the bulk modulus ($B_v$), shear modulus ($G_v$) and Young's modulus ($E_v$). **Fig. N5b** compares the polarization switching barriers obtained from CI-NEB calculations using DFT and those predicted by DP. The two methods exhibit nearly identical energy barriers. In addition, the phonon spectrum of the $P6_3mc$ phase calculated using the DP model agrees well with that obtained from density functional perturbation theory (DFPT), indicating that the DP model can accurately describe the mechanical behavior and lattice dynamics of the crystal (**Fig. N5c**). Overall, the trained DP model exhibits high accuracy and is well suited for large scale atomistic simulations of MD in $V_N$ −AlN.

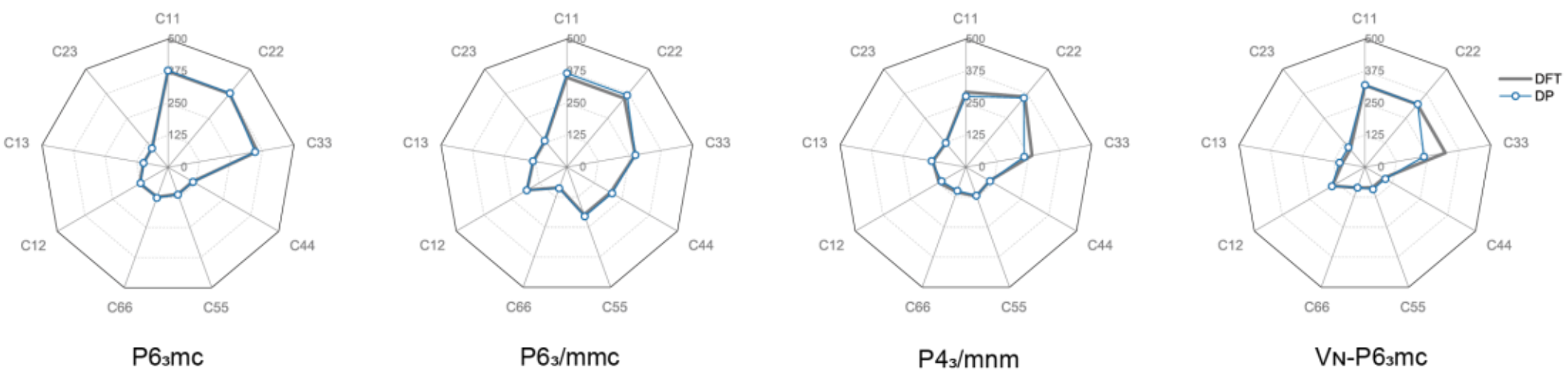

**Figure N4.** Elastic constants predicted by DP and calculated by DFT of pristine $P6_3mc$, $P6_3/mmc$, $\beta$ −BeO−like AlN and contain 1 $V_N$ $P6_3mc$ AlN

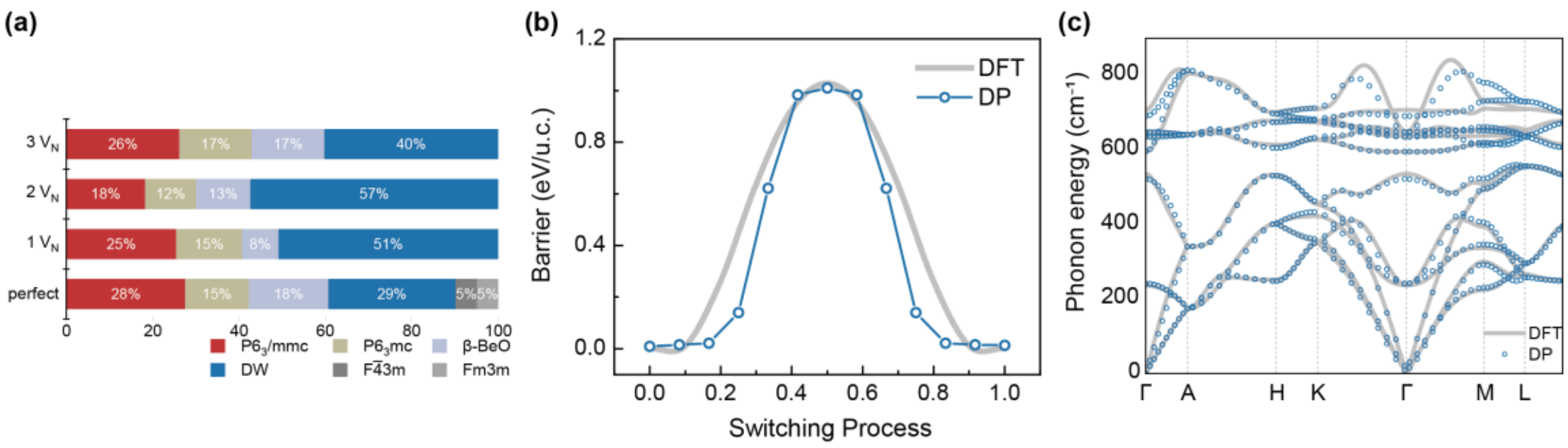

**Figure N5. a,** Proportion of configuration types in the dataset used to train the DP model. **b,** Comparison of DW switching barriers predicted by the DP model and calculated by DFT. **c,** comparison of the phonon spectra of the $P6_3mc$ phase obtained using DP and DFT calculations.

Table N1. Elastic constants, $B_v$, $G_v$ and $E_v$ for $P6_3mc$, $P6_3/mmc$ and $\beta$ −BeO-like phases

| | | $P6_3mc$ | $P6_3/mmc$ | $\beta$ −BeO−like |
|---|---|---|---|---|
| $C_{11}$ | DP | 377.421 | 366.4702 | 276.626 |
| | DFT | 374.4865 | 350.743 | 293.106 |
| | **error** | **0.78%** | **4.48%** | **5.62%** |
| $C_{22}$ | DP | 377.421 | 366.4702 | 352.9787 |
| | DFT | 374.4865 | 350.743 | 358.268 |
| | **error** | **0.78%** | **4.48%** | **1.48%** |
| $C_{33}$ | DP | 345.9192 | 272.1735 | 231.4613 |
| | DFT | 354.6529 | 271.722 | 263.082 |
| | **error** | **2.46%** | **0.17%** | **12.02%** |
| $C_{44}$ | DP | 113.3085 | 204.503 | 109.0571 |
| | DFT | 111.8762 | 198.528 | 107.723 |
| | **error** | **1.28%** | **3.01%** | **1.24%** |
| $C_{55}$ | DP | 113.3085 | 204.5028 | 118.704 |
| | DFT | 111.8762 | 198.528 | 121.021 |
| | **error** | **1.28%** | **3.01%** | **1.91%** |
| $C_{66}$ | DP | 126.9467 | 87.95713 | 98.72977 |
| | DFT | 123.3734 | 85.363 | 104.238 |
| | **error** | **2.90%** | **3.04%** | **5.28%** |
| $C_{12}$ | DP | 123.5277 | 181.3278 | 110.2271 |
| | DFT | 127.7397 | 180.016 | 121.486 |
| | **error** | **3.30%** | **0.73%** | **9.27%** |
| $C_{13}$ | DP | 96.66689 | 134.5632 | 136.3609 |
| | DFT | 97.3169 | 133.13 | 130.869 |
| | **error** | **0.67%** | **1.08%** | **4.20%** |
| $C_{23}$ | DP | 96.66685 | 134.5632 | 122.7077 |
| | DFT | 97.3169 | 133.13 | 122.331 |
| | **error** | **0.67%** | **1.08%** | **0.31%** |
| $B_v$ | DP | 208.1588 | 243.0419 | 165.6934 |
| | DFT | 209.9886 | 236.925 | 178.6927 |
| | **error** | **0.87%** | **2.58%** | **7.27%** |
| $G_v$ | DP | 118.7637 | 159.7303 | 98.71402 |
| | DFT | 116.4751 | 153.2622 | 98.9578 |
| | **error** | **1.96%** | **4.22%** | **0.25%** |
| $E_v$ | DP | 299.3588 | 393.0787 | 247.0758 |
| | DFT | 294.9007 | 378.23 | 250.6115 |
| | **error** | **1.51%** | **3.93%** | **1.41%** |

## Note 2. First Principles Calculations

### Born effective charges

To simulate the effect of an electric field in MD, we defined the atomic charges from the Born effective charge tensor. The Born effective charges were calculated with the Berry-phase method, as follows:

$$Z^* = -\frac{P_{dis} - P_{per}}{d}$$

here, $P_{dis}$, $P_{per}$ and $d$ denote the dipole moment after perturbation of the selected atom along $z$, the dipole moment before perturbation, and the displacement distance, respectively.

### Normalized trace of interatomic force constants (IFC) tensors

To evaluate the change in bond strength around the vacancy, we calculated the normalized trace of the IFC tensors. The IFC tensors were obtained from density functional perturbation theory and are defined as:

$$\frac{\partial^2 E}{\partial R_i R_j} = \begin{pmatrix} \frac{\partial^2 E}{\partial R_x R_x} & \frac{\partial^2 E}{\partial R_x R_y} & \frac{\partial^2 E}{\partial R_x R_z} \\ \frac{\partial^2 E}{\partial R_y R_x} & \frac{\partial^2 E}{\partial R_y R_y} & \frac{\partial^2 E}{\partial R_y R_z} \\ \frac{\partial^2 E}{\partial R_z R_x} & \frac{\partial^2 E}{\partial R_x R_y} & \frac{\partial^2 E}{\partial R_z R_z} \end{pmatrix}$$

where $E$ denote the total energy and $R_\alpha$, $\alpha \in (x, y, z)$ denote the atomic position along a given direction. To compare atomic interactions between the defect containing and intrinsic systems, we further calculated the normalized IFC trace as:

$$normalized\ trace\ of\ IFC = \frac{\frac{\partial^2 E}{\partial R_{1,x} R_{2,x}} + \frac{\partial^2 E}{\partial R_{1,y} R_{2,y}} + \frac{\partial^2 E}{\partial R_{1,z} R_{2,z}}}{\frac{\partial^2 E}{\partial R_{1,x} R_{1,x}} + \frac{\partial^2 E}{\partial R_{1,y} R_{1,y}} + \frac{\partial^2 E}{\partial R_{1,z} R_{1,z}}}$$

where $\frac{\partial^2 E}{\partial R_{1,x} R_{2,x}}$ denotes the second order force constant between atoms $a$ and $b$ along the $x-$direction, while $\frac{\partial^2 E}{\partial R_{1,x} R_{1,x}}$ represents the self-interaction of atom $a$ along the $x-$ direction. By normalizing the IFCs, the interatomic interactions in different compounds and bonding environments can be quantitatively compared.

### Vacancy formation energy

The defect formation energy was calculated as follows:

$$E_{format} = E_{defect} - E_{perfect} + \mu + q(E_V + E_F + \Delta V)$$

where $E_{defect}$ denotes the internal energy of the defective system, $E_{perfect}$ denotes the internal energy of the perfect crystal, $\mu$ is the chemical potential of the defect species, $q$ is the net charge of the system, $E_V$ represents the valence band maximum, $E_F$ is the Fermi−level position, and $\Delta V$ is the correction term accounting for the valence band alignment between the defective and perfect crystals.

For Sc-doped AlN systems, under N-rich conditions, the chemical potential of nitrogen, $\mu_N$, was taken as one half of the total energy of an $N_2$ molecule. Under Al-rich conditions, $\mu_N$ was obtained from $\mu_N = \frac{\Delta H}{n} - \mu_{Al}$, where $\Delta H$ is the enthalpy of the system, $n$ is the number of atoms in the system, and $\mu_{Al}$ is the chemical potential of Al, taken as the per-atom energy of elemental Al.

## Note 3. Molecular Dynamics simulations

A tetragonal supercell of size $20 \times 25 \times 10$ was employed, corresponding to a bulk block of approximately $10.8\ nm \times 7.8\ nm \times 5.2\ nm$ and containing 40,000 atoms. After equilibration in an NPT ensemble at 300 K for 10 $ps$, the system was subjected to electric field control. During all simulations, the $x-$ and $y-$directions were constrained, whereas the $z-$direction was allowed to relax to mimic the stress effect imposed by the electrodes.

In MD simulations, polarization switching may still occur at longer times even when it is not observed within a short timescale under the applied electric field, however, MD simulations are inherently limited in accessing long time evolution. To balance computational cost, two schemes of driving electric fields were employed in this study. The nucleation distributions were obtained from 20 $ps$ MD simulations under linearly increasing electric fields driving polarization switching. All point hysteresis loops were simulated using staged, linearly increasing electric fields (as shown in **Fig. N6**), with each stage lasting 10 $ps$.

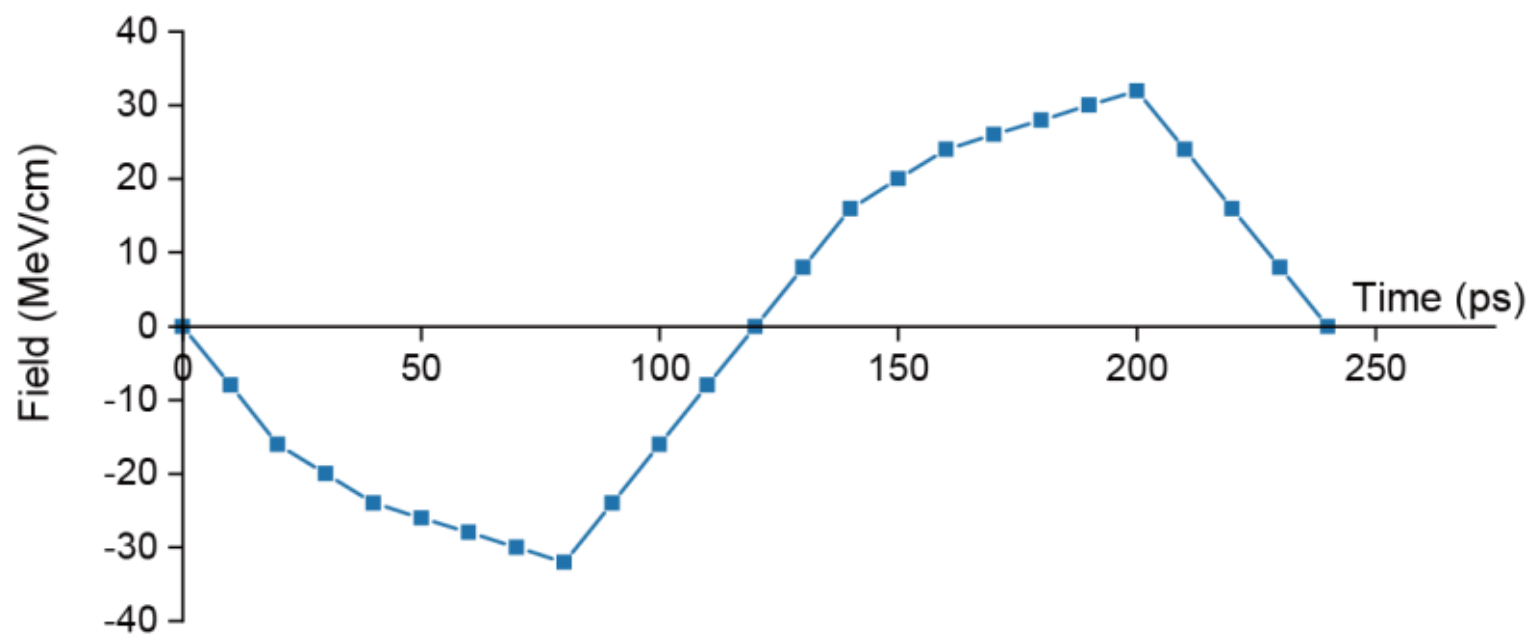

**Figure N6.** Electric field applied for simulating the hysteresis loop.

## Supplementary Figures

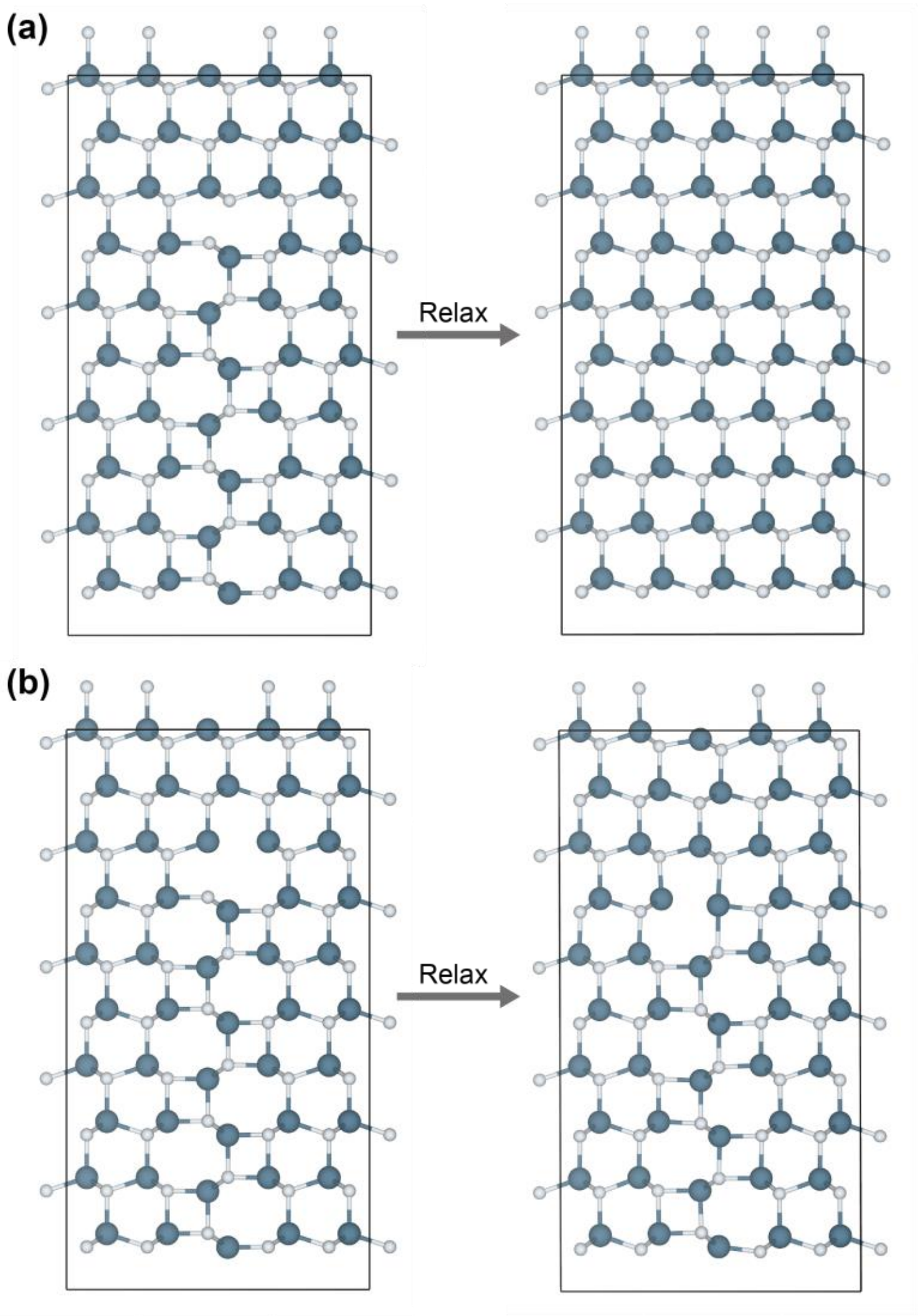

**Figure 1. a,** Crystal structures of the $5 \times 5 \times 1$ pristine AlN supercell containing an introduced DW before and after structural relaxation. **b,** Crystal structures of the AlN supercell containing one $V_N$ with an introduced DW before and after structural relaxation.

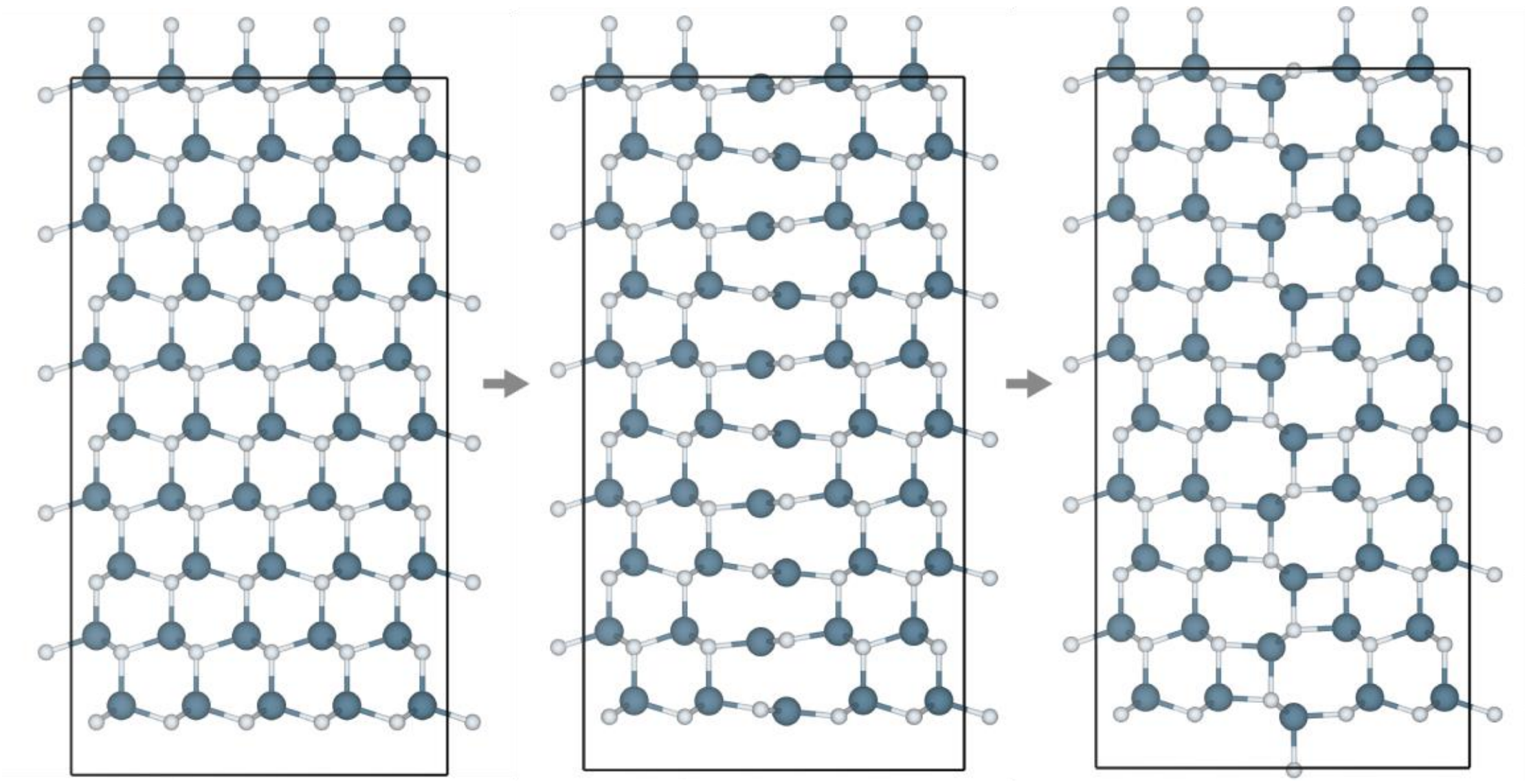

**Figure 2.** Intermediate structures along the CI-NEB pathway for polarization switching in the $5 \times 5 \times 1$ pristine AlN supercell.

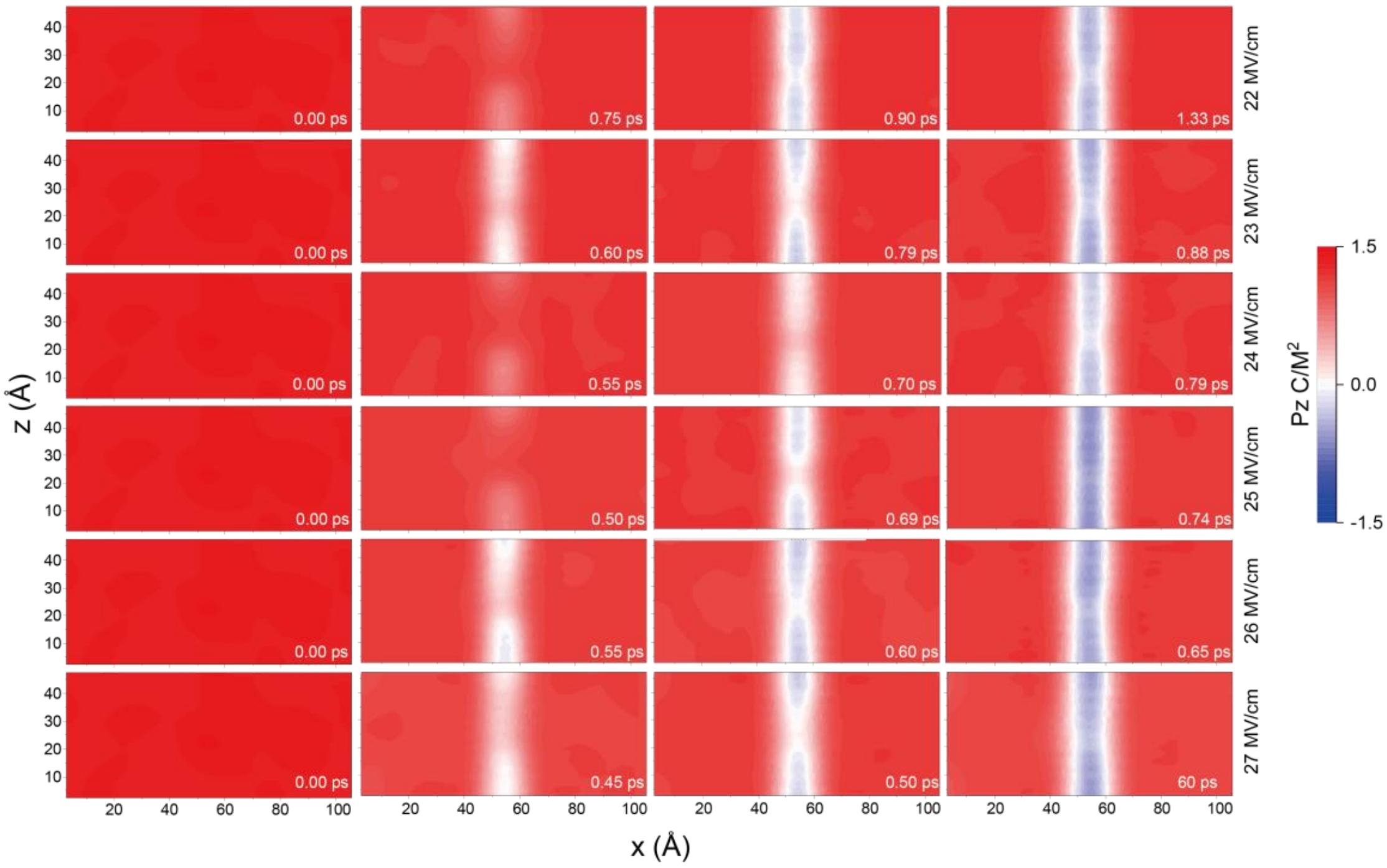

**Figure 3.** Polarization distribution during domain nucleation near the $V_N$ cluster under electric fields ranging from $22$ to $27\ MV/cm$.

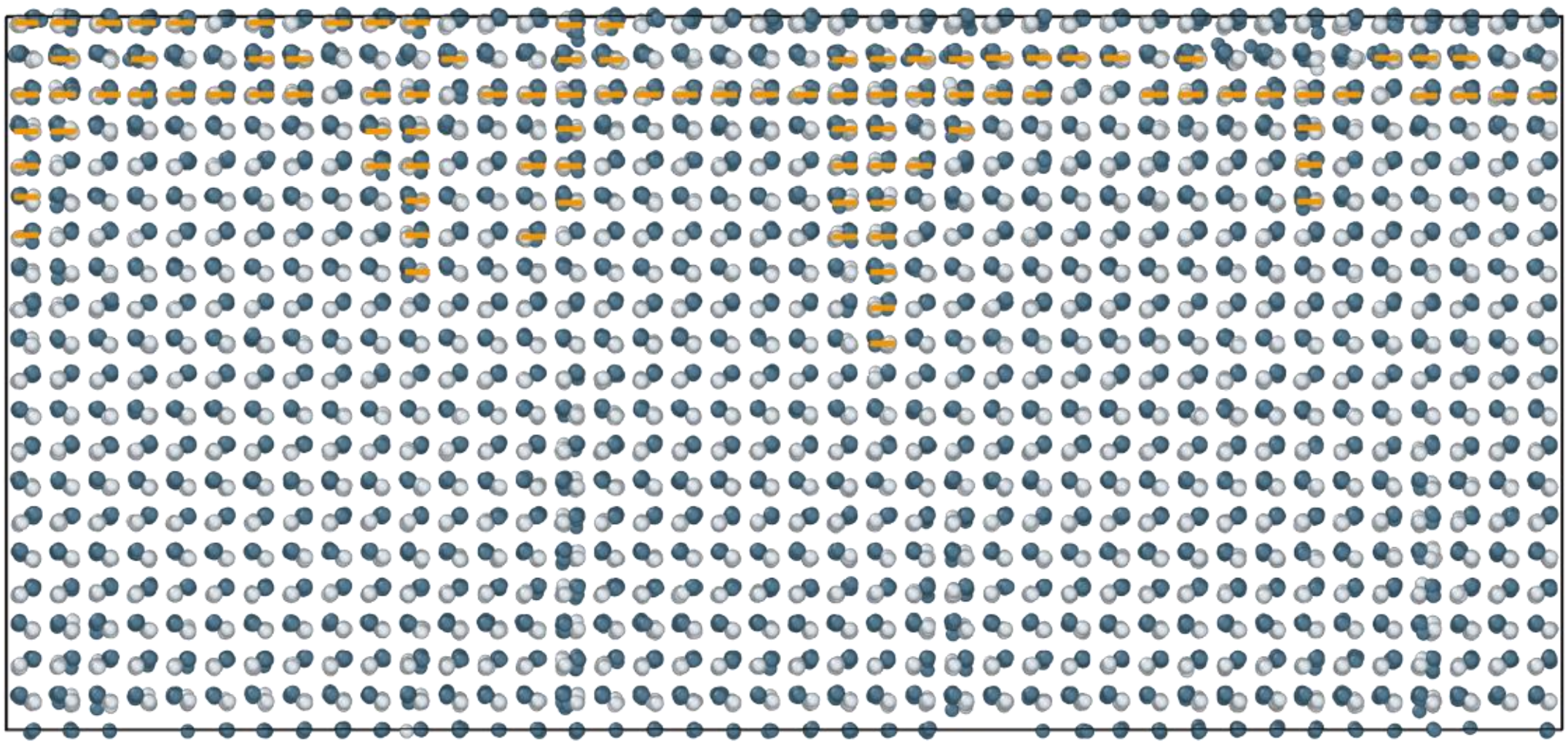

**Figure 4.** Domain distributions projected onto the xz plane obtained from MD simulations of the $V_N$-rich layer model under an electric field of $21\ MV/cm$. Blue and white spheres represent Al and N atoms, respectively, while orange regions indicate the reversed domains formed near the $V_N$-rich layer.

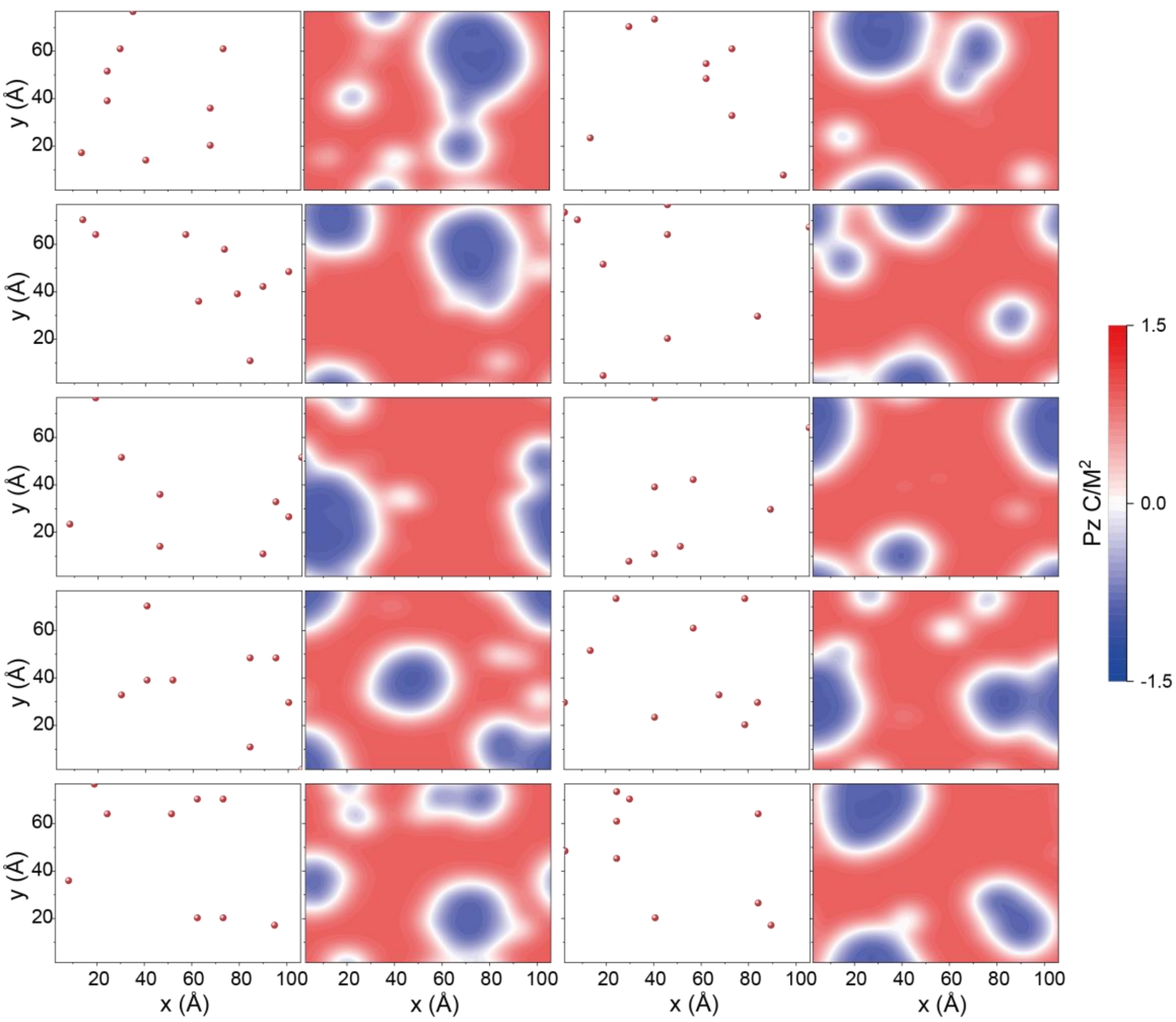

y (Å)
20
40
60
80
100
x (Å)
Pz C/M2
1.5
0.0
-1.5

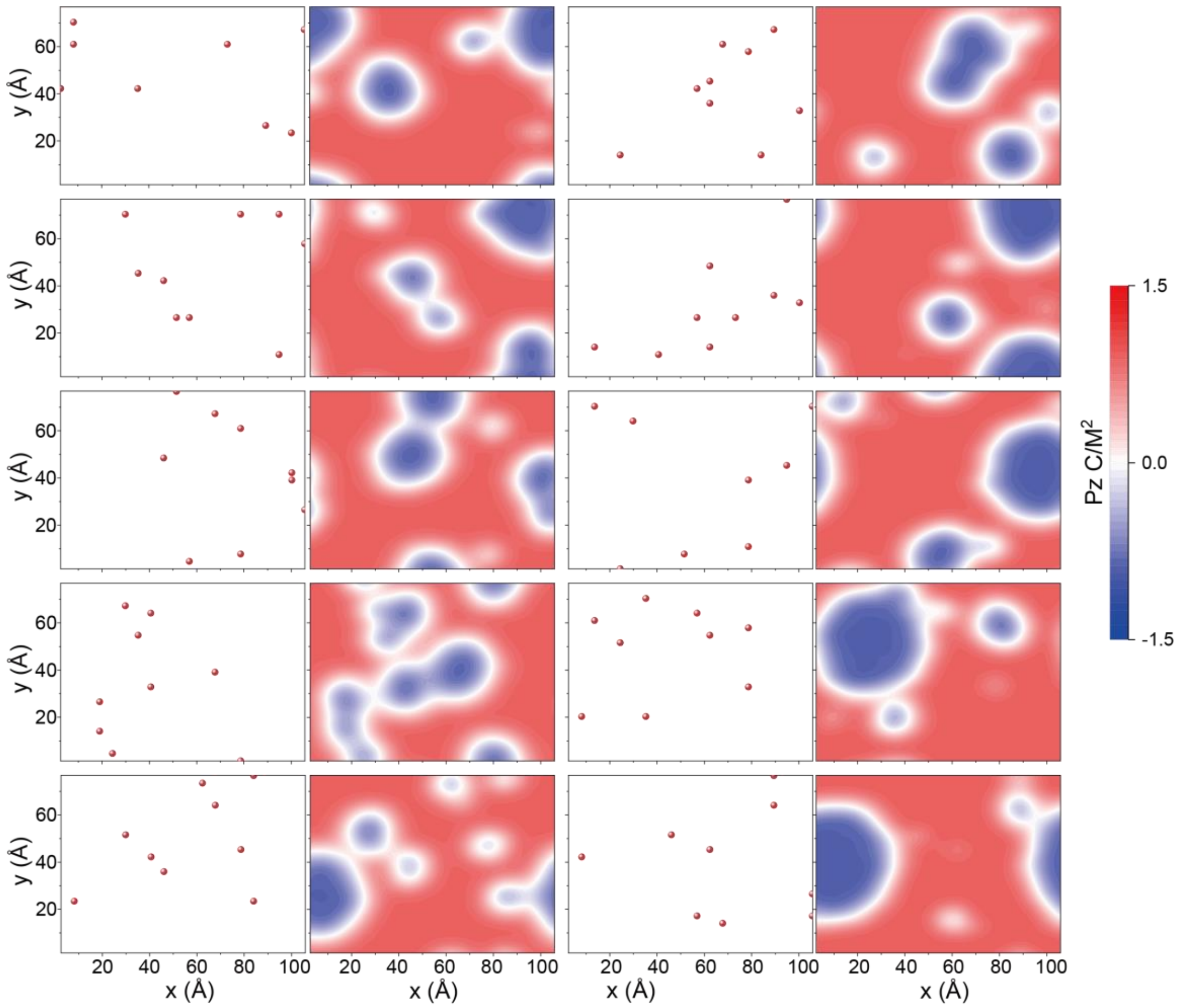

**Figure 5.** Multiple simulations performed with randomly distributed N vacancies under an electric field of $21\ MV/cm$ to exclude stochastic artifacts.

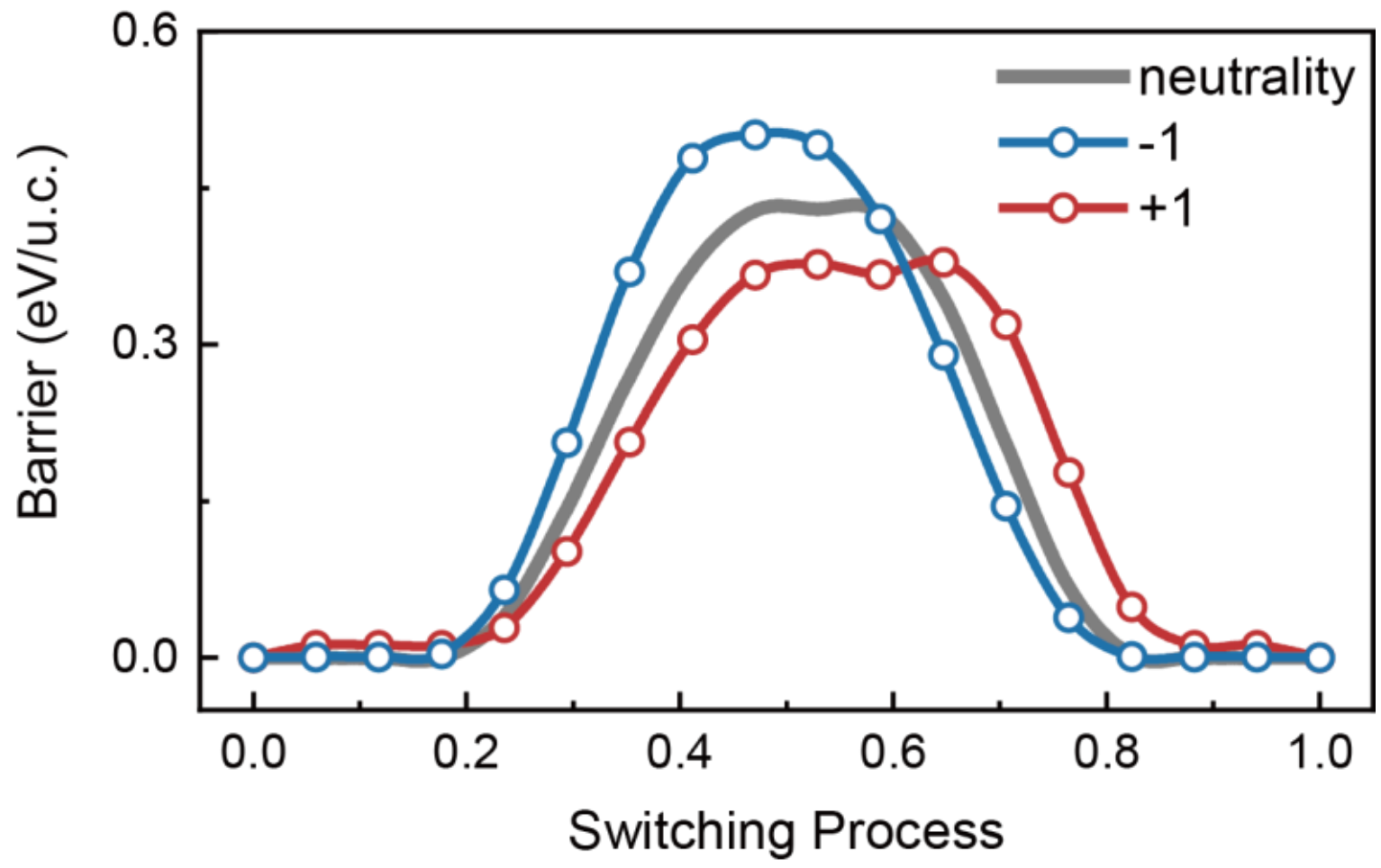

**Figure 6.** CI-NEB barriers for polarization switching in charged AlN containing $V_N$.

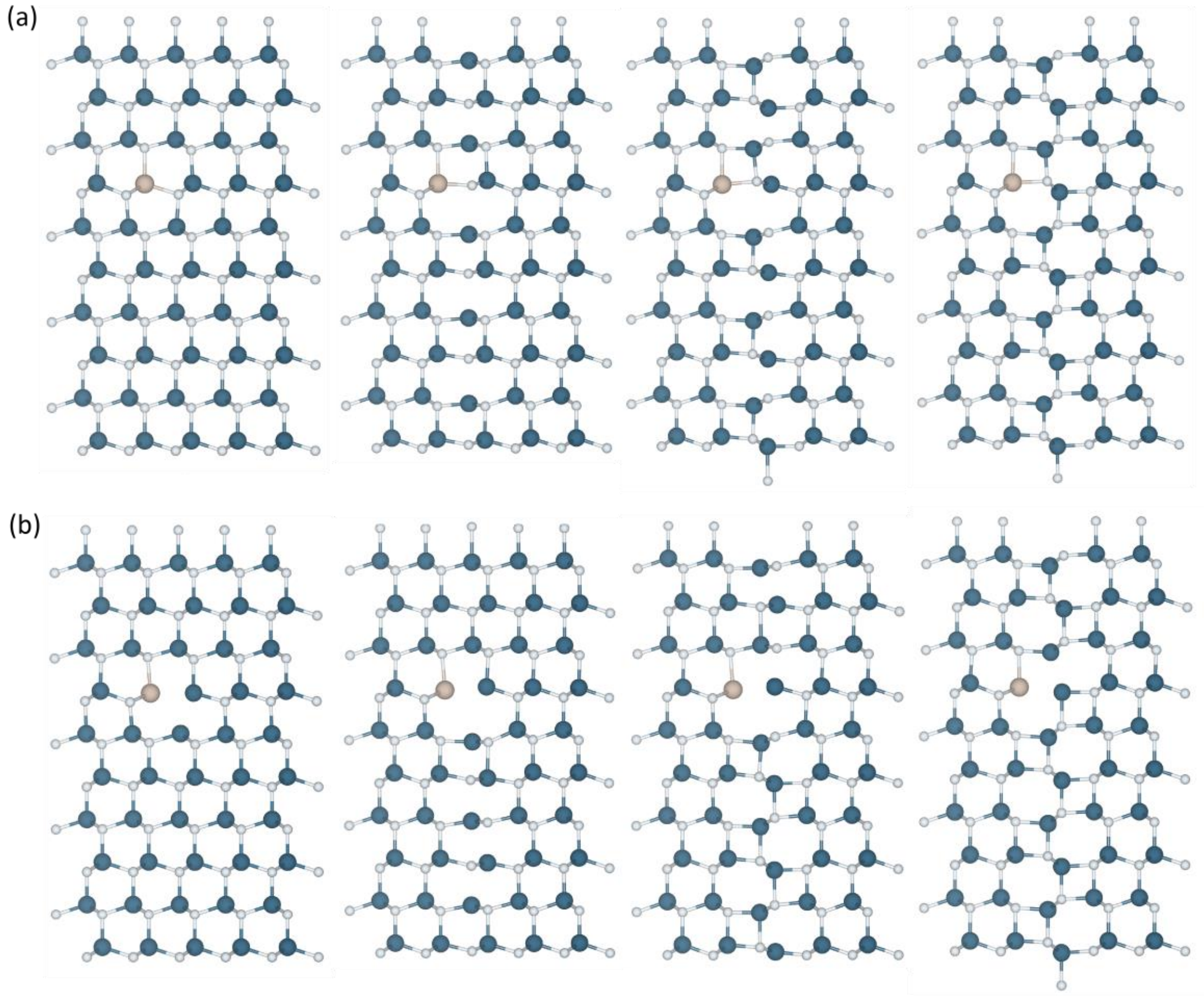

**Figure7.** CI-NEB images of single-Sc-doped AlN(a) defect-free structure and (b) structure containing a $V_N$.